\documentclass[floatfix,aps,prd,twocolumn,a4paper,showpacs,amssymb,amsmath,amsfonts,nofootinbib,longbibliography]{revtex4-1}
\usepackage[utf8]{inputenc}
\usepackage{xspace}
\usepackage{acronym}
\usepackage{booktabs}
\usepackage[table,dvipsnames]{xcolor}
\usepackage[colorlinks, citecolor=ForestGreen, linkcolor=blue, pdfborder={0 0 0}, breaklinks=true]{hyperref}
\usepackage[caption=false]{subfig}

\usepackage{tabularx}
\usepackage{multirow}
\usepackage{makecell}

\usepackage{scrextend}
\usepackage{textgreek}

\usepackage{aas_macros}
\usepackage{dcolumn}
\usepackage{graphicx,longtable,mathrsfs,color,array}
\usepackage{placeins}
\usepackage{bm}
\usepackage{siunitx}
\usepackage{bold-extra}
\usepackage{soul}
\usepackage{lipsum}

\newcommand{\hypo}{\mathcal{H}_s}

\begin{document}

\title{Overlapping signals in next-generation gravitational wave observatories: \\
A recipe for selecting the best parameter estimation technique}

\author{Tomasz Baka$^{1, 2}$}
\author{Harsh Narola$^{1, 2}$}
\author{Justin Janquart$^{3, 4}$}
\author{Anuradha Samajdar$^{1,2}$}
\author{Tim Dietrich$^{5, 6}$}
\author{Chris Van Den Broeck$^{1, 2}$}

\affiliation{${}^1$Institute for Gravitational and Subatomic Physics (GRASP), Utrecht University, 3584 CC Utrecht, The Netherlands}
\affiliation{${}^2$Nikhef, 1098 XG Amsterdam, The Netherlands}
\affiliation{${}^3$Center for Cosmology, Particle Physics and Phenomenology - CP3, Universit\'e Catholique de Louvain, Louvain-La-Neuve, B-1348, Belgium}
\affiliation{${}^4$Royal Observatory of Belgium, Avenue Circulaire, 3, 1180 Uccle, Belgium}
\affiliation{${}^5$ Institut f\"{u}r Physik und Astronomie, Universit\"{a}t Potsdam, Haus 28, Karl-Liebknecht-Str. 24/25, 14476, Potsdam, Germany}
\affiliation{${}^6$ Max Planck Institute for Gravitational Physics (Albert Einstein Institute), Am M\"uhlenberg 1, Potsdam, Germany}

\date{\today}

\begin{abstract}
Third-generation gravitational wave detectors such as Einstein Telescope and 
Cosmic Explorer will have significantly better sensitivities than current 
detectors, as well as a wider frequency bandwidth. This will increase the number and duration of 
the observed signals, leading to many signals overlapping in time. If not adequately accounted for, this 
can lead to biases in parameter estimation. In this work, we combine 
the joint parameter estimation method with relative binning to handle full parameter
inference on overlapping signals from binary black holes, including precession effects and 
higher-order mode content. As this method is computationally more expensive than traditional 
single-signal parameter estimation, we test a prior-informed Fisher matrix and a 
time-frequency overlap method for estimating expected bias to help us decide when joint parameter 
estimation is necessary over the simpler methods. We improve upon previous Fisher matrix
implementations by including the prior information and performing an optimization routine to better locate
the maximum likelihood point point, but we still find the method unreliable. The time-frequency method
is accurate in 86\% of close binary black hole mergers. We end 
by developing our own method of estimating bias due overlaps, where we reweight the single signal parameter estimation
posterior to quantify how much the overlapping signals affect it. We show it has 99\% accuracy for zero noise injections (98\% in
Gaussian noise), at the cost of one additional standard sampling run when joint parameter estimation proves to be necessary.
\end{abstract}
\maketitle

\section{Introduction}

The continuous upgrade of the Advanced LIGO~\cite{LIGOScientific:2014pky} and 
Advanced Virgo~\cite{VIRGO:2014yos} detectors over the last ten years has made the observation of 
gravitational waves (GWs) routine, with 
already more than 200 GW candidate detections reported during the O4 observing runs~\cite{gracedb}. 
The in-depth analysis of the events observed during the previous observing runs~\cite{KAGRA:2021vkt} 
has already enabled major advances in fundamental physics~\cite{LIGOScientific:2021sio}, 
astrophysics~\cite{LIGOScientific:2021psn}, and cosmology~\cite{LIGOScientific:2021aug}, 
and the increase in number of observed GW signals should further improve the precision of the measurements. Even further upgrades to these second-generation detectors~\cite{a_sharp_upgrades}, as well as the addition of other detectors to the network---such as KAGRA~\cite{Somiya:2011np,Aso:2013eba} and LIGO-India~\cite{ligo-india-1}---will further increase the number of observed signals and the precision of the measurements.
Additionally, a next generation of ground-based detectors is planned, with Einstein Telescope 
(ET)~\cite{Punturo:2010zz,et2025} in Europe and Cosmic Explorer (CE)~\cite{Reitze:2019iox} 
in the United States; these are referred to as third-generation (3G) observatories. They are expected to 
lead to a steep change in GW observations, 
both in terms of sensitivity and increased frequency bandwidth~\cite{satyhya2012}. As a result, 
it will become common for signals to overlap in 
time~\cite{Regimbau:2009rk,Samajdar:2021egv,Pizzati:2021apa,Relton:2021cax,Himemoto:2021ukb}. 

Studies have shown that overlapping compact binary coalescence (CBC) signals can lead to biased results
in the measurement of source parameters if not properly accounted 
for~\cite{Samajdar:2021egv,Pizzati:2021apa,Relton:2021cax,Janquart:2022fzz}. In turn, this  
can negatively impact the inference of CBC mass distributions or tests of 
general relativity~\cite{Hu:2022bji}. Therefore, it is important to develop methods to properly 
account for overlapping signals. A first step in that direction was made in~\citet{Janquart:2022fzz}, 
where we have applied joint parameter estimation (JPE) and hierarchical subtraction (HS) to 
a population of high-mass binary black hole (BBH) signals. There, we showed that JPE can accurately 
recover the parameters of the individual signals, while HS works in some cases but often leads to 
biased results for close merger times. However, the JPE method is computationally expensive, which is a severe 
inconvenience when accounting for the large detection rate expected in the future. Therefore, we 
argued that a good strategy would be to apply HS by default and JPE whenever one knows that bias 
is present if the overlap is neglected. However, \emph{a priori} there is no obvious way of knowing when 
to expect significant bias. An approach that has been proposed in the literature is to use 
appropriately adapted Fisher matrices to predict bias that would occur if overlaps are 
neglected~\cite{Antonelli:2021vwg}. Here, we build upon this and more recently developed Fisher matrix 
analyses~\cite{Dupletsa:2024gfl} to also include prior information in the result.
We find that the method is reliable for large biases, but that it fails to accurately predict the shift 
in the posterior probability distributions when they are comparable to the effect of the detector 
noise. In contrast, we demonstrate that a time-frequency overlap method~\cite{johnson2024} 
shows greater reliability in predicting the bias.

Here we propose a different approach where we perform parameter estimation (PE) on a 
single signal and reweight the posterior to one with the best estimate of the other signal 
removed from the strain. By calculating the Jensen-Shannon divergence (JSD) between the original 
and the reweighted posterior we are able to accurately predict if the overlapping signals affected PE. 
If they did not, we already have the accurate posterior. If they did, we perform JPE to 
fully account for the overlapping signals.

To make the application of JPE to realistic 3G-like data computationally feasible,
we rely on a modification of the PE framework developed in~\citet{Janquart:2022fzz}, 
which now includes relative binning (RB)~\cite{Zackay:2018qdy,Dai:2018dca,Leslie:2021ssu,Narola:2023men}. 
With this, we are able to cope with scenarios including precession, higher-order modes (HOMs), 
and a lower analysis cutoff frequency of $5\,\mathrm{Hz}$. 

This work is structured as follows. In Sec.~\ref{sec:pe} we outline the different 
PE methods and demonstrate JPE with RB produces accurate posteriors. In 
Sec.~\ref{sec:bias_estimation}, we show the performance of different methods of estimating  
bias due to signal overlaps, and we introduce and assess our new PE-based methodology.  
We close with our conclusions in Sec.~\ref{sec:conclusions}.

\section{\label{sec:pe}Parameter estimation methods}
The goal of GW parameter estimation for CBCs is to fully characterize the parameters $\vec{\theta}$ of the 
binary---such as chirp mass $\mathcal{M}_c$, mass ratio $q$, companion spins $a_1,a_2$---by obtaining 
their posterior probability density function (PDF): $p(\vec{\theta} | d, \hypo)$, conditioned on the 
observed data $d$ and the hypothesis $\hypo$ that a signal is present in the data and it is accurately 
described by the model we use. To avoid clutter, we will drop $\hypo$ from further equations, but it 
should be understood that every probability is conditioned on it. From Bayes' theorem, we get an expression for the posterior~\cite{Veitch:2009hd}:
\begin{equation}\label{eq:BayesTheorem}
    p(\vec{\theta} | d) = \frac{\mathcal{L}(d | \vec{\theta})p(\vec{\theta})}{\mathcal{Z}(d)} \, ,
\end{equation}
with $\mathcal{L}(d | \vec{\theta})$ being the likelihood of parameters given data, $p(\vec{\theta})$ is the prior probability of parameters, and $\mathcal{Z}(d)$ is the normalization factor called evidence, given by:

\begin{equation}\label{eq:evidence}
    \mathcal{Z}(d) = \int \mathrm{d}\vec{\theta}\, p(d | \vec{\theta}) p(\vec{\theta}) \, .
\end{equation}

For GW parameter estimation, $d(t)$ is the time domain strain data recorded by the interferometer network, split between the signal component $h(t)$ and the noise component $n(t)$:
\begin{equation}\label{eq:strain}
    d(t) = n(t) + h(t) \,.
\end{equation}
When we consider overlapping signals, $h(t)$ can be split further between individual signal contributions. In this paper, we limit ourselves to overlaps of two signals, denoted by $A$ and $B$. With this decomposition, we can rewrite the data strain as:

\begin{equation}\label{eq:TwoSignalsData}
    d(t) = n(t) + h_{A}(t) + h_B(t) \,.
\end{equation}

Assuming colored Gaussian noise, the likelihood takes the 
form:
\begin{equation}\label{eq:SingleEventLikelihood}
    \log\mathcal{L}(d | \vec{\theta}) \propto  -\frac{1}{2} \langle d - h(\vec{\theta}) | d - h(\vec{\theta}) \rangle  \, ,
\end{equation}
with the inner product $\langle \cdot | \cdot \rangle$ defined as

\begin{equation}\label{eq:InnerProduct}
    \langle a | b \rangle = 4 \Re \int_{f_{\rm{low}}}^{f_{\rm{high}}}  \mathrm{d}f\, \frac{a^{*}(f)b(f)}{S_{n}(f)} \,.
\end{equation}
Here $a(f), b(f)$ are Fourier transforms of time domain functions $a(t), b(t)$; $^*$ denotes the complex 
conjugate; $f_{\rm{low}}$ and $f_{\rm{high}}$ are the limiting frequencies of the analysis, and 
$S_n(f)$ is the one-sided power spectral density (PSD).

\subsection{Single-signal parameter estimation}
The simplest way to handle the actual PE of overlapping signals is to ignore the presence of a second signal, assuming a template for a unique merger as in Eq.~\eqref{eq:SingleEventLikelihood}. Generally, we recover only the parameters of the loudest signal present in the data at that time, possibly with biased estimates due to the other underlying signals. These biases can potentially be sufficiently large that the true parameters are outside of the recovered distribution. If the signals are close in SNR, we may recover the parameters of the quieter signal instead. For injections performed in this paper, we have never observed the sampler mixing both signals in the posterior.
Sampling with \textsc{Bilby} uses differential evolution~\cite{2006S&C....16..239T} proposal method for drawing new nested sampling live points. Two overlapping signals have their modes far enough away in the parameter space that the sampler becomes fixated on one of the modes and does not propose jumps between them.

\subsection{Hierarchical subtraction}
In hierarchical subtraction, we improve on the previous method by sequentially performing 
single signal parameter estimation (SSPE) analyses and subtracting best-fitting signals. First, we run the SSPE on the full data containing both signals, $d(A,B)$. We recover the parameters of one of the 
signals, calling it $A$ (in general, this is the signal with higher SNR): $p(\vec{\theta}_A|d(A,B))$. 
From the posterior, we can get a point estimate of the parameters of the signal $\hat{\theta}_A$ 
(such as the maximum likelihood sample), and the associated waveform:
\begin{equation}\label{eq:BestFitA}
    \hat{h}_A(t) = h(t, \hat{ \theta}_A).
\end{equation}
We can then subtract this waveform from the data to obtain:

\begin{align}\label{eq:SubtractedData}
    d(B, r_A, t) &=  d(A, B, t) - \hat{h}_A(t) \nonumber\\
    &= n(t) +h_B(t)+h_A(t)-\hat{h}_A(t) \nonumber \\
    &= n(t) +h_B(t)+r_A(t) \, ,
\end{align}
with $r_A(t) = h_A(t)-\hat{h}_A(t)$ the residual due to imperfect recovery of signal $A$. 
If the process works correctly, this residual should have an SNR much lower than that of signal $B$.

We can now run SSPE on the reduced data $d(B, r_A, t)$ to obtain the estimates of 
the posterior of signal $B$, $p(\vec{\theta}_A|d(B, r_A))$, and corresponding best point estimate 
$\hat{\theta}_B$.

We can repeat the steps above, this time subtracting signal $B$ and performing PE on:
\begin{align}
    d(A, r_B, t) &=  d(A, B, t) - \hat{h}_B(t)
\end{align}
to get a new estimate of the posterior of the signal $A$. In principle, we can continue 
this process iteratively however long we wish.

This approach is simple to implement, as it requires only SSPE runs and it was shown to produce good results in many cases~\cite{Janquart:2022fzz}. However, it is not guaranteed to converge to the correct posterior. After subtracting one of the signals, the residual remaining in the strain can still be strong enough to bias PE. Additionally, it cannot account for possible correlations in the posteriors between the signals,  as during each PE step the parameters of one of the signals are effectively fixed.

\subsection{Joint parameter estimation}
The methods described above all use approximations to simplify the analysis, but we can also model all 
the signals merging in the data (in the cases considered in this paper there are only 2). 
We modify the signal template as:
\begin{equation}
    h(\vec{\theta}) = h_A(\vec{\theta}_A) + h_B(\vec{\theta}_B),
\end{equation}
with $h_A,h_B$ the individual signal templates. With this, we can modify the likelihood 
to~\cite{Janquart:2022fzz}:
\begin{align}\label{eq:Likelihood2signals}
    & \mathcal{L}(d(A, B) | \vec{\theta})\nonumber \\ 
    &\propto \exp \Bigg[ -\frac{1}{2} \bigg\langle d(A, B) - h_A(\vec{\theta}_A) - h_B(\vec{\theta}_B) \bigg| \nonumber \\ 
    & \,\,\,\,\,\,\,\,\,\,\,\,\,\,\,\,\,\,\,\,\,\,\,\,\,\,\,\,\,\,\,\,\,\,\,\,\,\,\,\,\,\,d(A, B) - h_A(\vec{\theta}_A) - h_B(\vec{\theta}_B) \bigg\rangle  \Bigg] \, ,
\end{align}
with the expanded parameter space $\vec{\theta}=(\vec{\theta}_A,\vec{\theta}_B)$.

Of note is that the above likelihood is symmetric under the exchange of signals $A$ and $B$, 
so the posterior will inherit this symmetry.
Still, for PE it is useful to break this symmetry between 
signals completely, to accurately recover the properties of each signal. 
To this end, we sort the posteriors to enforce an ordering of chirp masses $\mathcal{M}_c$ to follow 
the ordering in the injection. 
We check if $\mathcal{M}_{\mathrm{inj},A} > \mathcal{M}_{\mathrm{inj},B}$, then for each sample $\theta_i$ we compute $(\mathcal{M}_{i,A} > \mathcal{M}_{i,B})=(\mathcal{M}_{\mathrm{inj},A} > \mathcal{M}_{\mathrm{inj},B})$. If true, we keep the sample as is; if false, we swap each parameter A with the corresponding parameter B for this sample.
This works for signals considered in this work, but for signals with very close chirp masses, 
some other method of ordering the parameters would need to be employed.

While providing the most accurate method of modeling overlapping signals, this 
JPE method suffers greatly from computational constraints. 
The likelihood takes twice as long to compute (as we have to generate two waveforms for each 
likelihood call), but it is the expansion of parameter space that poses the biggest problem. 
For standard BBH analysis of a single signal, the parameter space is 15-dimensional. 
When doubled, the parameter space grows to 30 dimensions which is much higher than usual analysis 
and slows down parameter estimation considerably. In~\citet{Janquart:2022fzz}, it took 
24 days to complete JPE on a 16-core \textsc{Intel Skylake Gold 6148} CPU for high-mass overlapping signals and a starting frequency of 
20 Hz. With 3G detectors, signals will be longer due to lower starting frequency, and, consequently, PE will be slower. To tackle JPE in a realistic scenario, we need to combine it with 
a method of speeding up the likelihood evaluation.

\subsection{Relative binning}
In general, the frequency domain GW strain in the detector, $h(\vec{\theta}, f)$ is a highly 
oscillatory function, which in turn requires high sampling frequency to accurately capture its 
behavior. However, we do not need to capture the waveform accurately 
over all the parameter space $\vec{\theta}$, but just in the region where 
the likelihood $\mathcal{L}(\vec{\theta}|d)$ is close to its maximum value. 
Relative binning (RB) is a method of likelihood approximation which exploits this, 
approximating the ratio of a waveform to a reference waveform by a piecewise linear 
function~\cite{Zackay:2018qdy,Leslie:2021ssu,Narola:2023men}. RB is a form of the heterodyne principle, which was first proposed for use in GW data analysis in~\citet{2010arXiv1007.4820C}.

Let us denote by $\vec{\theta}'$ the reference waveform parameters, and by $h'(f)$ the corresponding 
reference waveform. $\vec{\theta}'$ is chosen to be close to the maximum of $\mathcal{L}(\vec{\theta}|d)$. 
While $h(f)$ varies quickly with frequency $f$ of the signal, the ratio $h(f)/h'(f)$ varies 
much more slowly if we remain close to the maximum likelihood point, requiring fewer frequency points 
to capture the signal behavior. Let us denote by $b = [f_{min}, f_{max}]$ a particular frequency bin 
in which we evaluate the signal, whose width is chosen such that the waveform ratio 
can be approximated as
\begin{equation}
    \frac{h(f)}{h'(f)} = r_1(b)+r_2(b)(f-f_c(b)) +\mathcal{O}(f^2) \,,
\end{equation}
and the quadratic correction is negligible. $f_c(b)$ is the central frequency of the bin and $r_i(b)$ coefficients can be calculated from the waveform ratio at the bin edges.

This binning scheme performs well for nonprecessing signals with just the (2,2) mode 
present~\cite{Zackay:2018qdy} and for this case it is implemented in the \textsc{Bilby} 
package~\cite{Krishna:2023bug}. With higher modes present, the ratio $h(f)/h'(f)$ still 
varies quickly with frequency and many bins are required to capture the signal. 
To improve RB, we can apply it to each mode of the signal individually, so that 
we will not have to worry about capturing the beating pattern between the modes~\cite{Leslie:2021ssu,Narola:2023men}.
The detector strain can be written as:
\begin{equation}
    h(f)=\sum_{l,m}(F_+C^+_{l,m}(f) + F_\times C^\times_{l,m}(f))h^L_{l,m}(f)\,.
\end{equation}
Here $(l, m)$ denote component modes of the signal; $(F_+, F_\times)$ are the detector 
response functions; $C^+_{l,m}(f), C^\times_{l,m}(f)$ are the twisting-up functions, 
transforming the signal from the co-precessing frame $L$ to the total angular momentum frame; and  
$h^L_{l,m}(f)$ are the individual modes of the signal in the co-precessing frame. 
We define $C_{l,m}(f) = F_+C^+_{l,m}(f) + F_\times C^\times_{l,m}(f)$ and perform RB mode-by-mode:
\begin{align}
    \frac{h_{l,m}(f)}{h'_{l,m}(f)} &= r^1_{l,m}(b)+r^2_{l,m}(b)(f-f_c(b)) +\mathcal{O}(f^2)\\
    \frac{C_{l,m}(f)}{C'_{l,m}(f)} &= s^1_{l,m}(b)+s^2_{l,m}(b)(f-f_c(b)) +\mathcal{O}(f^2)\,,
\end{align}
where $C'_{l,m}(f)$ is the twisting-up function for the reference waveform. $r^i_{l,m}(b)$ and $s^i_{l,m}(b)$ by the ratio of waveform modes and twisting-up functions at the bin edges.
With the above 
approximation, the inner products in the likelihood function can be approximated by:
\begin{widetext}
\begin{alignat}{3}
    \langle d | h \rangle &= \Re \sum_b \sum_{(l,m)}&&\left[  A^0_{(l,m)}(b) r^{0*}_{(l,m)}(b) s^{0*}_{(l,m)}(b) +A^1_{(l,m)}(b) \left(  r^{1*}_{(l,m)}(b) s^{0*}_{(l,m)}(b) + r^{0*}_{(l,m)}(b) s^{1*}_{(l,m)}(b) \right)   \right], \nonumber\\
    \langle h | h \rangle &= \Re \sum_b \sum_{(l,m)} \sum_{(\bar{l},\bar{m})} &&\left\{ B^0_{(l,m),(\bar{l},\bar{m})}(b) r^0_{(l,m)} r^{0*}_{(\bar{l},\bar{m})} s^0_{(l,m)} s^{0*}_{(\bar{l},\bar{m})} \right.\nonumber\\
    & &&+ B^1_{(l,m),(\bar{l},\bar{m})}(b) \left[ s^0_{(l,m)} s^{0*}_{(\bar{l},\bar{m})} \left(  r^0_{(l,m)} r^{1*}_{(\bar{l},\bar{m})} + r^{0*}_{(\bar{l},\bar{m})} r^1_{(l,m)}\right)\right.\nonumber\\
    & &&+ \left.\left.r^0_{(l,m)} r^{0*}_{(\bar{l},\bar{m})} \left(  s^0_{(l,m)} s^{1*}_{(\bar{l},\bar{m})} + s^{0*}_{(\bar{l},\bar{m})} s^1_{(l,m)} \right)\right]\right\}, \label{eq:binned_dot_product}
\end{alignat}
\end{widetext}
with coefficients $A^0, A^1, B^0, B^1$ computed once, before the sampling starts, and defined by:
\begin{align}
    A^0_{(l,m)}(b) &= \frac{4}{T} \sum_{f \in b} \frac{d(f) h_{(l,m)}^{\prime L*}(f)}{S_n(f)} , \nonumber\\ 
    A^1_{(l,m)}(b) &= \frac{4}{T} \sum_{f \in b} \frac{d(f) h_{(l,m)}^{\prime L*}(f)}{S_n(f)}\left(f-f_c(b)\right) , \nonumber \\
    B^0_{(l,m), (\bar{l}, \bar{m})}(b) &= \frac{4}{T} \sum_{f \in b} \frac{h_{(l,m)}^{\prime L}(f) h_{(\bar{l},\bar{m})}^{\prime L*}(f)}{S_n(f)} , \nonumber \\
    B^1_{(l,m), (\bar{l}, \bar{m})}(b) &= \frac{4}{T} \sum_{f \in b} \frac{h_{(l,m)}^{\prime L}(f) h_{(\bar{l},\bar{m})}^{\prime L*}(f)}{S_n(f)} \left(f-f_c(b)\right) .
\end{align}
We then choose the frequency bins $b$ for our signal by choosing a fiducial waveform different 
from the reference waveform and ensuring that the difference between the true log-likelihood and 
the approximate log-likelihood for this waveform falls below the desired error tolerance, 0.01 for our paper.
An example of this approach is proposed in~\citet{Leslie:2021ssu}, where the full frequency range is 
recursively bisected until the given error tolerance is reached and we use the same approach here.

We use and build upon the computationally efficient implementation in~\cite{Narola:2023men,janquart:2022} 
where the authors created a fast, 
\textsc{Bilby}-compatible~\cite{Ashton:2018jfp,Romero-Shaw:2020owr} 
implementation for the \textsc{IMRPhenomXPHM} waveform model~\cite{Pratten:2020ceb}.

\subsection{Combining joint parameter estimation with relative binning}
\begin{figure}
    \centering
    \includegraphics[width=0.9\columnwidth]{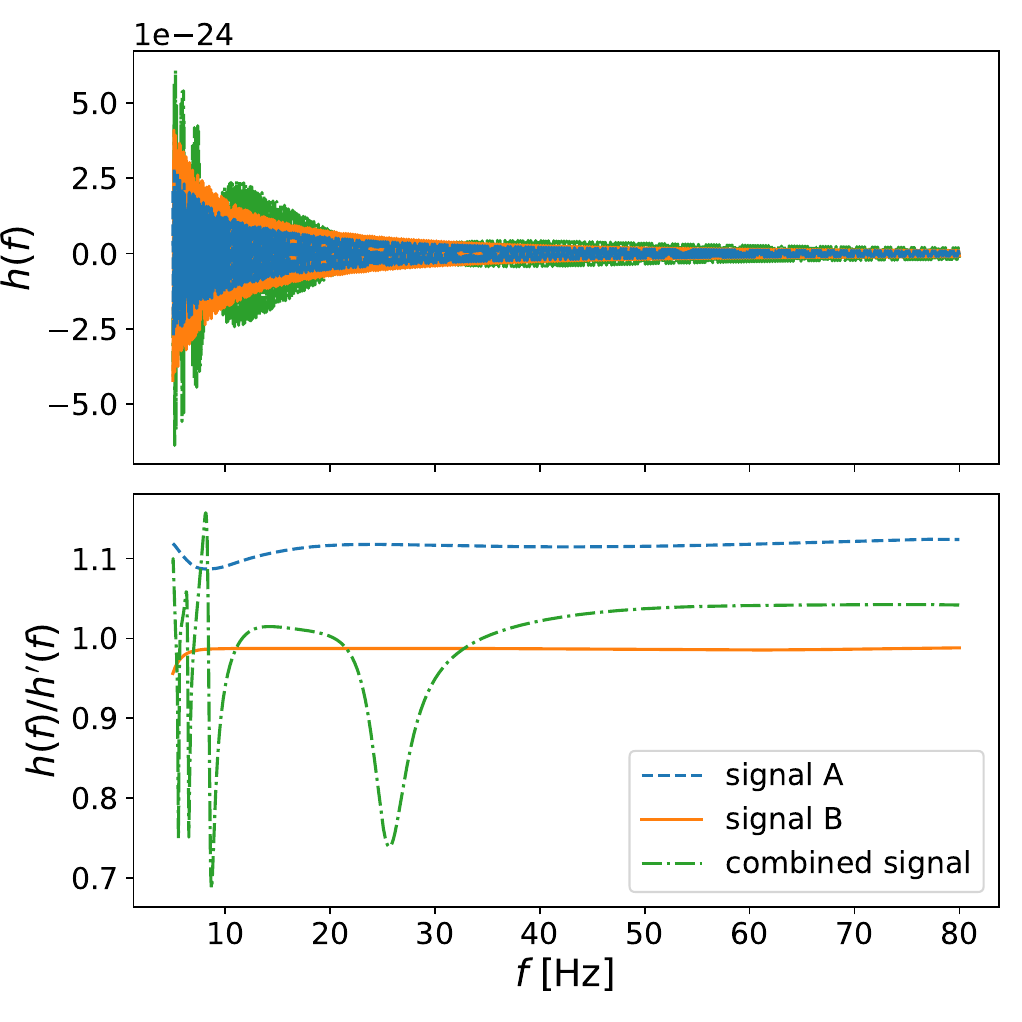}
    \caption{RB is applied to two overlapping signals. They were chosen to be non-precessing and to contain just the dominant (2,2) mode, for clarity of the figure. \emph{Top:} Frequency domain strain of the individual and the combined signals. They are strongly oscillating, requiring high sampling frequency to be captured properly. 
    \emph{Bottom:} Real part of the ratio of the strain to the reference strain. The ratio of the combined signal to the combined reference signal is much smoother than the absolute signal. The ratios of the composite signals are much smoother and are better suited for approximation by a piecewise linear function than the combined ratio.}
    \label{fig:strain_ratios}
\end{figure}

The advantages of using RB for likelihood evaluations are that it is very fast and, for 
our use case, that it easily generalizes to multiple signals in the detector. Indeed, the separate 
spin-weighted harmonics already behave like separate signals, so the only change necessary is to 
change the indexing in Eq.~~\eqref{eq:binned_dot_product} from $(l, m)$ to $(i, l,m)$, where $i$ labels 
the signals present in the 
detector. We must mention that while this works fine for two 
signals, it does scale quadratically with the number of signals present:  
the number of waveform inner products we need to calculate to obtain the likelihood scales 
as $N(N+1)/2$ where $N$ is the number of signals. The number of frequency bins needed to 
capture the signal stays constant, as we bin each signal individually.

An example of how RB works for multiple signals is shown in Fig.~\ref{fig:strain_ratios}. At the 
top, we plot the real part of the frequency domain strain for the individual and the combined 
signals. We see that it is strongly oscillating and as such requires high resolution 
to arrive at a proper representation. In the bottom plot, we show the ratio of the strain 
to the reference strain. If we consider the ratios of the combined signals, the function is much 
smoother than the absolute strain and requires fewer points to represent it as a piecewise linear function. But if we choose to represent each signal ratio individually, the functions are even less oscillating, leading to just a few frequency points needed for the likelihood approximation.

For this paper, we modify the mode-by-mode RB implementation from~\cite{Narola:2023men} to be compatible 
with multiple signals and use it throughout to produce our results. This implementation is only compatible with the \textsc{IMRPhenomXPHM} waveform model, as twisting-up procedure is waveform dependent. For completeness we additionally implemented 
JPE with RB for a generic waveform model without precession nor higher modes. 
Details about the code release can be found in Appendix~\ref{app:code_release}.

\subsection{\label{sec:jpe_test}Tests of joint parameter estimation with relative binning}
For our tests of JPE with RB, we use a detector network comprised of one 10 km triangular ET observatory located at the Virgo site\footnote{The design of ET has not been finalized between two L-shaped detectors and one triangular detector, the latter of which we chose for this work. For comparisons between designs, see~\citet{2023JCAP...07..068B}.} and two CEs  
located at sites of LIGO-Hanford and LIGO-Livingston having arm lengths of 40 km and 20 km respectively. 
We inject two BBH signals with the \textsc{IMRPhenomXPHM} waveform 
model~\cite{Pratten:2020ceb} into Gaussian noise following projected noise 
curves for ET~\cite{Hild:2010id} and CE~\cite{LIGOScientific:2016wof}. We 
set the lower frequency cutoff of the analysis at 5 Hz, for which we note that 
traditional PE approaches become untractable for large injection studies. 
We then analyze the signals with the \textsc{Bilby}~\cite{Ashton:2018jfp,Romero-Shaw:2020owr} 
software using \textsc{Dynesty}~\cite{Speagle2019Dynesty:Evidences} as nested sampler. 
We use our custom implementation of the RB likelihood~\footnote{\url{https://github.com/corvus314/jointRB}}.

We draw 80 signal pairs from the priors summarized in Table~\ref{tab:pp_prior} and inject 
them into Gaussian noise. We only use injections for which SNRs are above 8. 
We keep merger times of both signals close to each other to test JPE on cases where overlapping 
signal bias is expected to be significant~\cite{Pizzati:2021apa}.

\begin{table}
    \begin{tabular}{ l  l }
    \hline
    \textbf{Parameter} & \textbf{Prior} \\
    \hline
    $\mathcal{M}_c$ &  UniformInComponents($10\,M_{\odot},$ $100\,M_{\odot}$) \\
    $q$ & UniformInComponents(0.1, 1) \\
   $a_{1,2}$  & Uniform(0, 1) \\
    $\theta_{1,2}$ &Sin($0, \pi$)\\
     $\Delta \phi, \phi_{JL}$ & Uniform($0, 2\pi$)\\
    RA &  Uniform($0, 2\pi$)\\
    DEC &  Cos($-\pi/2,$ $\pi/2$)\\
   $ \theta_{JN}$ &  Sin($0, \pi$)\\
    $\psi$ &  Uniform($0, \pi$)\\
   $\phi_c$ &  Uniform($0, 2\pi$)\\
    $D_L$ &  UniformSourceFrame(0.1 \text{Gpc}, 40 \text{Gpc}) \\
    $t_c$ &  Uniform(-0.05 s, 0.05 s) \\
    \hline
     \end{tabular}
    \caption{
        Summary of generation and analysis priors for tests of JPE. Both overlapping signals share the same priors, with signal $A$ chosen to have higher SNR in the pair. 
        The parameters have their usual meanings: chirp mass $\mathcal{M}_c$, mass ratio $q$, 
        dimensionless spins $a_i$ together with their angles $\theta_i$ to the orbital 
        angular momentum, azimuthal angle between them $\Delta\phi$ and the precession angle 
        $\phi_{JL}$. RA and DEC stand for the right ascension and the declination 
        while $D_L$ is the luminosity distance to the source. 
        $\theta_{JN}$ is the angle between the line of sight and the total angular momentum 
        while $\psi$ is the polarization angle, while $\phi_c$ and $t_c$ stand for the phase 
        and time at coalescence. The UniformInComponents prior on $\mathcal{M}_c$ and $q$ 
        is equivalent to uniform priors in the component masses $m_1,\,m_2$ within 
        the prior boundaries. UniformSourceFrame is the distance prior uniform in comoving 
        volume and the source frame time.
    }
    \label{tab:pp_prior}
\end{table}

\subsubsection{Likelihood approximation}
We start with assessing how well our RB JPE implementation approximates the true likelihood with the two signals in the strain.
For constructing RB frequency bands, we follow~\cite{Narola:2023men} for the reference waveform and 
bin settings, using the injected signal for the former, and a change in 2\% for the chirp mass 
and mass ratio for the latter. 
We then recursively bisect the full frequency range until the RB error on the likelihood evaluated 
for the test waveform falls under a predetermined threshold, which we set to 0.03, following the default behavior of the RB code we adapted~\cite{janquart:2022}\footnote{To be more precise, we require that the likelihood error in each bin falls below $0.03/\sqrt{N_b}$, where $N_b$ is the proposed number of bins at each iteration of the bisection algortihm. We run the algorithm until the number of bins and the requested number of bins converge. After our analysis, we discovered that the RB implementation we adapted sometimes stops this process while one more bisection is possible, due to a rounding issue. This has led to excess likelihood error over the desired level. Even with this bug, our likelihoods were accurate enough that none of the posteriors produced in this work were affected.}.
These settings 
lead to a frequency grid of around a few hundred points, depending on the injected signal.
We note that for some injections (around 1 in 5) this leads to a frequency grid that 
does not capture the likelihood properly, but allowing for a different percentage difference 
(typically 10\%)
in chirp mass and mass ratio fixes the problem.

\begin{figure}
    \centering
    \includegraphics[width=0.9\columnwidth]{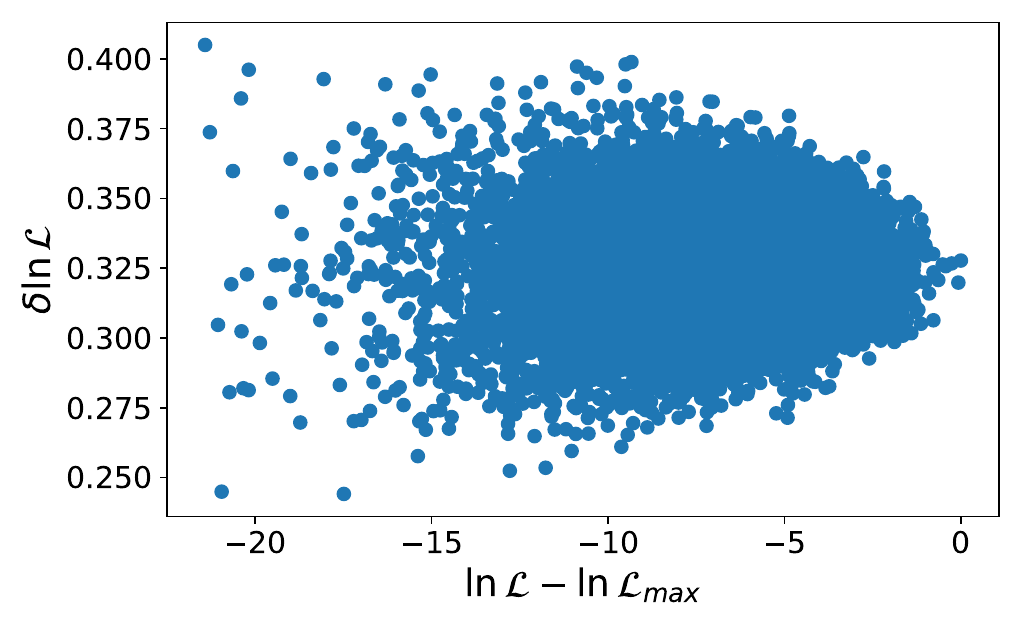}
    \caption{RB likelihood approximation error for an example overlapping signal. The scatter plot shows the distribution of likelihoods for all the points from the posterior. The x-axis shows the full unapproximated likelihood $\ln\mathcal{L}$ relative to the maximum likelihood sample $\ln\mathcal{L}_{max}$. The y-axis shows the absolute error $\delta\ln\mathcal{L}$ in the binned log-likelihood over the true likelihood. The relative error between the samples is at most 0.15, which leads to the effective sampling rate of $\eta_{eff}=99.96\%$.} 
    \label{fig:binning_error}
\end{figure}

In Fig.~\ref{fig:binning_error} we show how well the RB behaves for an example signal. We plot the RB log-likelihood error $\delta\ln\mathcal{L}=\ln\mathcal{L}_{RB}-\ln\mathcal{L}_{true}$ against the log-likelihood ratio $\ln\mathcal{L}$ of every sample from the posterior, relative to the maximum likelihood sample $\ln\mathcal{L}_{max}$. We see that the error on the likelihood clusters around the value 0.325, around 10 times larger than the error tolerance we have chosen. This can be explained by choosing the error tolerance not at any particular point from the posterior (which we do not know before sampling), but an arbitrary point in parameter space close to the injection, which is not the maximum likelihood point, due to presence of noise in the detector.

Even with $\delta\ln\mathcal{L} \approx 0.33$ all the likelihood errors lie close to this value, within 
the $(0.24, 0.41)$ range. We can show that this is enough to guarantee a good approximation of the 
likelihood.
Starting with samples drawn from the posteriors obtained with the RB likelihood, we can reweight them to the samples drawn from the true unapproximated likelihood by assigning to them weights equal to ratios of their likelihoods. We see from the definition of $\delta\ln\mathcal{L}$ that the expression for the weights is $w_i=\exp(-\delta\ln\mathcal{L}_i)$. From the weights, we can compute the effective sample size~\cite{elvira2022}:
\begin{equation}
    n_{eff} = \frac{(\sum_i^Nw_i)^2}{\sum_i^Nw_i^2} \,, 
\end{equation}

and the corresponding resampling rate $\eta_{eff} = n_{eff}/N$, which measures the fraction of the original $N$ samples we are effectively left with after the reweighting process. For our example signal, this is equal to $\eta_{eff} = 99.96\%$---high enough that the posterior obtained with RB is indistinguishable from the true posterior.\footnote{This assumes that the RB likelihood 
effectively covers the true posterior. If there are regions of the posterior for which RB likelihood 
erroneously sets the likelihood close enough to zero that we do not get any sample from them, 
$\eta_{eff}$ would be unaffected. If we instead got even a single sample from those regions, $\eta_{eff}$ 
would go to 0.}
For all signals investigated, the resampling rate is above 99.8\%, showing us that our implementation 
accurately approximates the true likelihood of overlapping signals and that any reweighting step is in 
practice unnecessary.

\subsubsection{Performance on test injections}
\begin{figure}
    \centering
    \includegraphics[width=0.9\columnwidth]{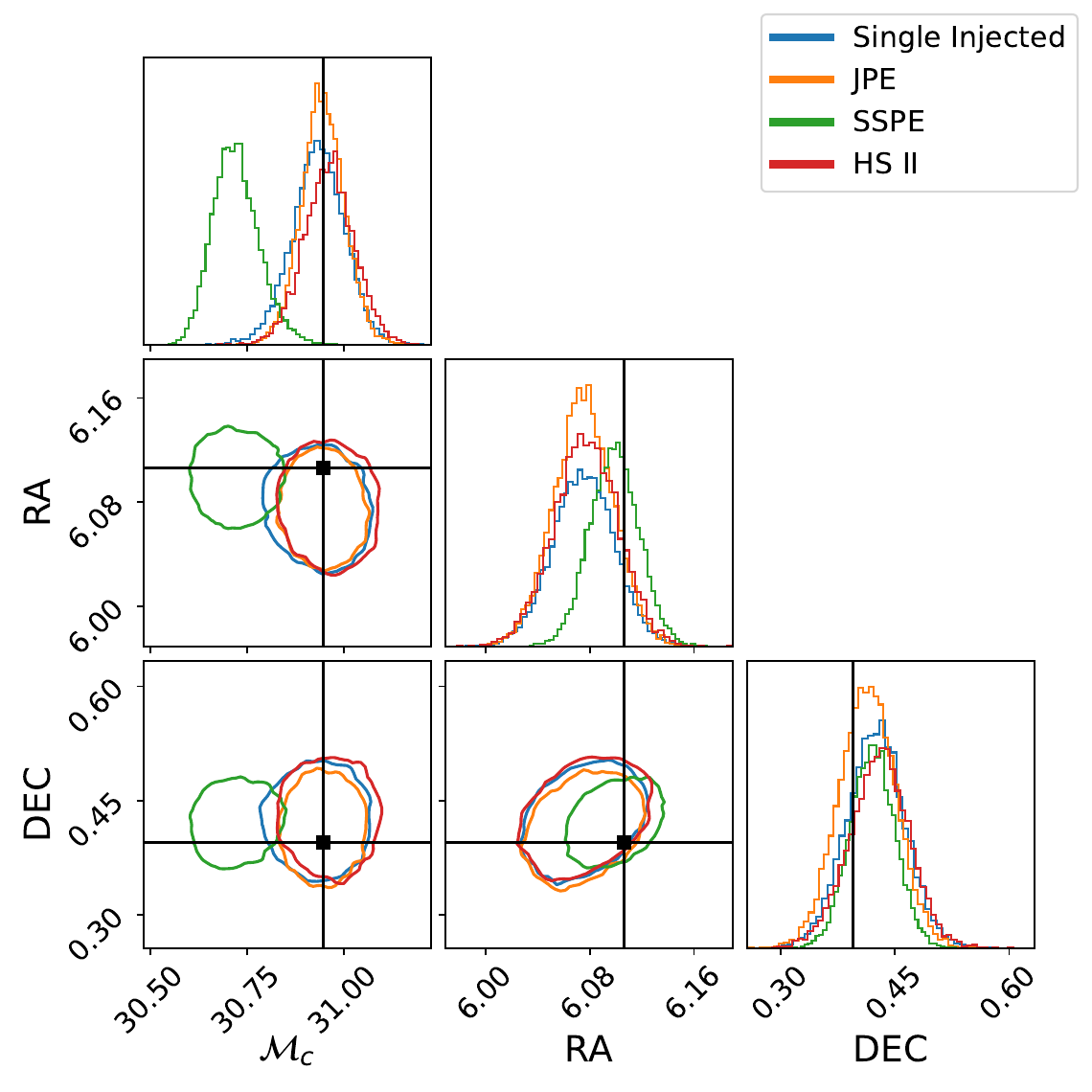}
    \caption{
        Comparison of different PE methods for overlapping signals and a representative signal in Gaussian noise. JPE is compared with SSPE and the second step of the HS method where the dominant signal is analized again after subtracting the estimation of signal B from the strain. For comparison, a posterior with only a single signal injected into the strain is also plotted. While both JPE and HS recover similar posteriors, they are statistically distinct. The contours in the corner plot indicate 0.86 credible regions.
    } 
    \label{fig:corner_normal}
\end{figure}

With the RB likelihood verified to accurately reflect the true likelihood of the two overlapping signals, we compare the performance of JPE against less computationally expensive methods, to demonstrate that some signals require modeling all the mergers present for accurate inference.

In Fig.~\ref{fig:corner_normal}, we show an example of a posterior with significant bias present. For many event pairs we analized, the presence of another signal in the strain has only a small effect on the posteriors. We have deliberately chosen to highlight one where the effect of the overlapping signal is very clear.

JPE posteriors represents what happens when we properly model all the signals present and as such is our ground truth, representing the true shape of the posterior, with the injected value properly recovered. It is very close to the posterior with no overlapping signal in the strain, a common feature for all 80 injections. In JPE, the uncertainty on the recovered value of $\vec{\theta}_B$ affects the recovery of $\vec{\theta}_A$, leading to a small difference between the two posteriors.

In contrast, when we do have overlapping signals present and try to recover just one, the resulting recovery tends to be biased. In this example, it is enough to shift the posteriors completely away from the injected value, but smaller shifts are common. In many cases, the biased posterior can be restored closer to its correct position by performing multiple hierarchical subtraction steps (HS II). In this example, subtracting the subdominant signal and reanalizing the data shifts the posterior close to the JPE posterior. There is still a small difference between the posteriors.

\begin{figure}
    \centering
    \includegraphics[width=0.9\columnwidth]{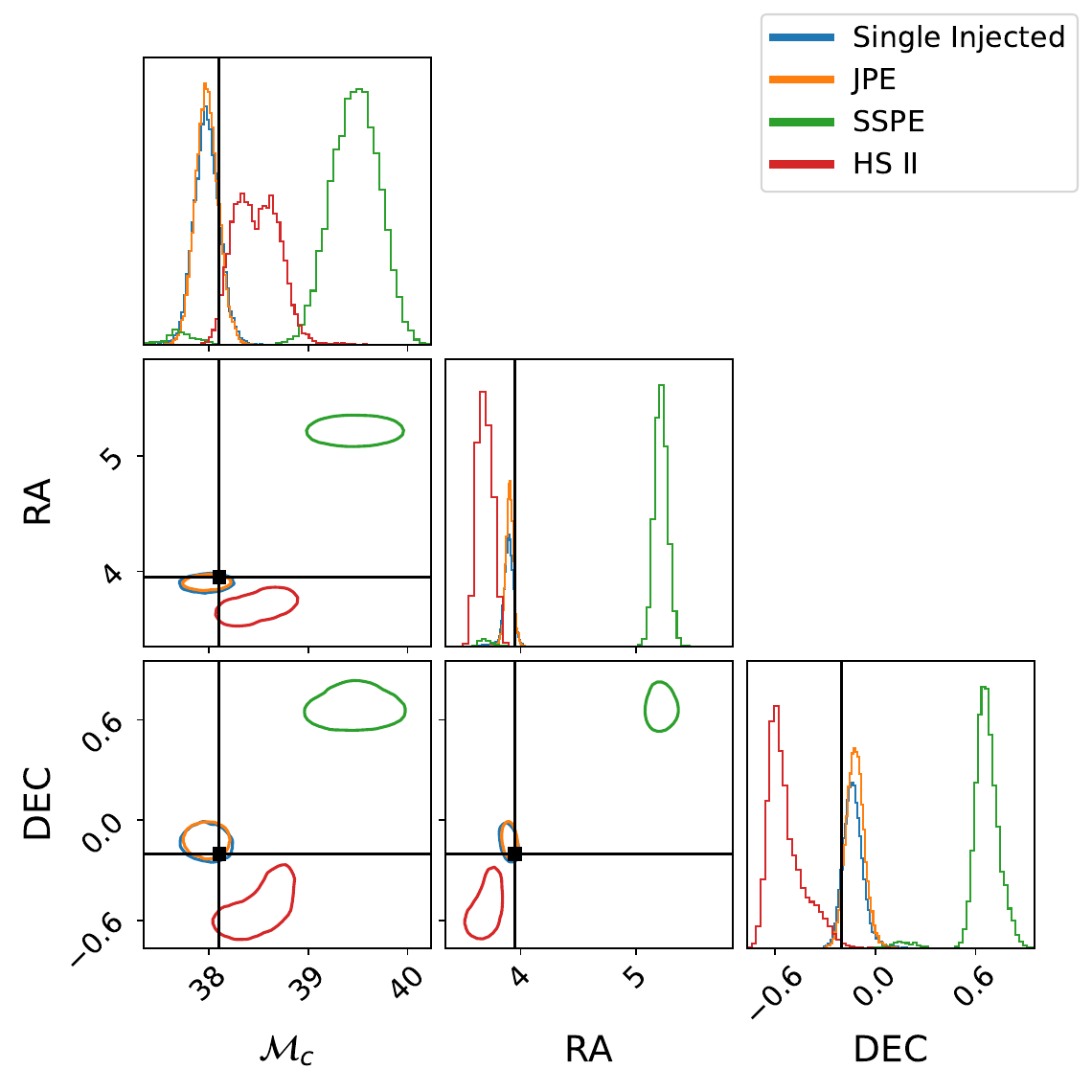}
    \caption{
        Comparison of different PE methods of overlapping signals for an extreme bias case. JPE is compared with SSPE and the second step of the HS method where the dominant signal is analized again after subtracting the estimation of signal B from the strain. For comparison, a posterior with only a single signal injected into the strain is also plotted. While both JPE and HS recover similar posteriors, they are statistically distinct. The contours in the corner plot indicate 0.86 credible regions. For this particular injection, SSPE case has a very small secondary peak closer to the injection, a feature very uncommon in the injections considered in this paper.
    } 
    \label{fig:corner_extreme}
\end{figure}

While this could suggest the less computationally expensive hierarchical subtraction could be used over JPE to get the posteriors quicker, we have to be careful as it does not perform well on all the events, like in the example posterior shown in Fig.~\ref{fig:corner_extreme}.
Here, when we recover overlapping signals with SSPE, the bias is more severe. While hierarchical subtraction reduces it substantially and shifts the posterior closer to JPE, it is not enough and the true values are still completely missed in the recovery. In principle, we could try more hierarchical subtraction steps to further improve the method, but more steps do not make sense from a computational point of view for two signals. With RB, JPE takes on average 11 days\footnote{For our set of injections the average SNR was 53.} on a 16-core CPU \textsc{Intel Skylake Gold 6148}, 6 times longer than a SSPE run and 2 times as long as the three PE runs needed for hierarchical subtraction. With two additional steps, most of the computational advantage of hierarchical subtraction would be lost (although others have attempted to speed up HS steps using neural networks~\cite{2025arXiv250705209H}). The method may still be useful where there are more than two signals overlapping in the strain.

\subsubsection{Percentile-percentile test}

To assess the accuracy of the overlapping signals posteriors when analyzing them with different PE methods, we perform a percentile-percentile test (pp test) of the posterior consistency~\cite{Veitch:2014wba,shaw2020}. We draw injections from the prior, perform parameter estimation with the same prior, get 1D distributions for each of the injection parameters, and record at which percentile in the posterior the injected value was recovered. We then plot the fraction of events that recovered the injected value in the given percentile. If the analysis was unbiased and the posterior samples were accurately drawn from the true posteriors, the percentiles should be distributed uniformly and the resulting graph should be a diagonal line (with some deviation due to the finite number of injections).

\begin{figure}
    \centering
    \includegraphics[width=0.9\columnwidth]{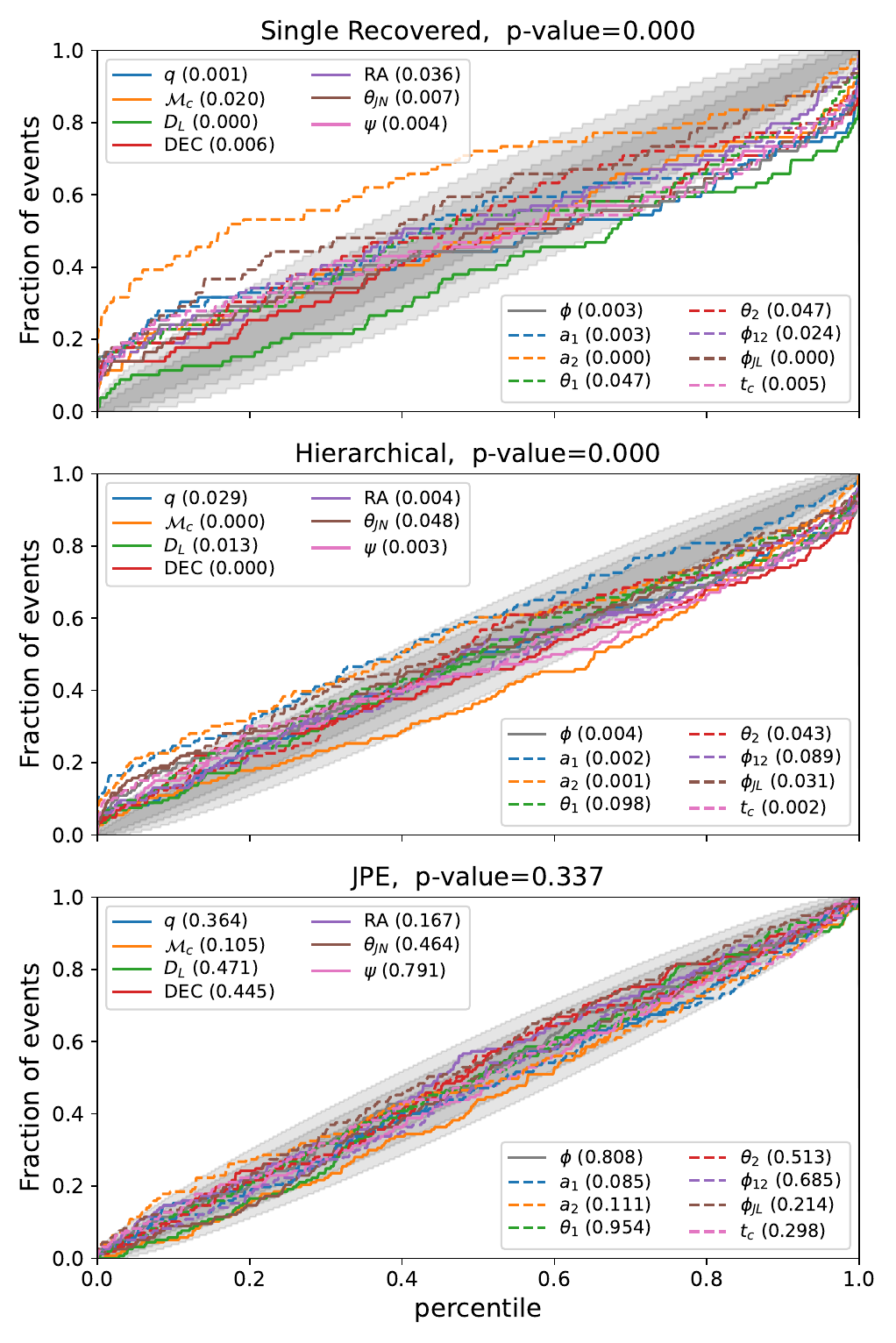}
    \caption{Percentile-percentile plot for different analysis methods of overlapping signals. The x-axis is the percentile at which the injection was recovered. The y-axis is the fraction of events for which the injection was recovered for a given percentile. The shaded regions indicate expected deviation from the diagonal at 0.68, 0.95, and 0.997 probability levels. The legend notes the p-values of the individual parameters, while the combined p-value is quoted above the plots. \emph{Top:} Recovery of the signal SSPE template. Only the louder signal can be recovered, resulting in 80 plotted injections. \emph{Middle:} Recovery of both signals using the hierarchical subtraction method, 160 plotted injections. \emph{Bottom:} Recovery of the signal with joint parameter estimation, 160 total signals.}
    \label{fig:pp_plot}
\end{figure}

In Fig.~\ref{fig:pp_plot}, we show how the different methods (all using RB approximation) recover the parameters of the signals. The cited p-values for individual parameters were computed using the Kolmogorov–Smirnov test (KS test)~\cite{ks-test} and combined using Fisher’s combined probability test~\cite{fisher1925statistical}.

At the top, we see how badly the standard PE method performed when applied to two overlapping signals. We can only recover the louder signal, resulting in only 80 injections plotted. The pp-plot significantly departs from a straight diagonal line, with a significant fraction of events recovered at 0.0 and 1.0 percentiles. For those events, the bias due to the presence of the second signal is strong enough that the injection lies completely outside the probability mass of the posterior. This is reflected by the very low p-values for all the parameters, and biases present in all the posteriors. The combined p-value is essentially equal to 0.

In the middle panel, we see how the hierarchical subtraction steps have improved the parameter recovery. After initial SSPE, we subtracted the recovered signal, estimated parameters of quieter signal, and estimated parameters of the louder signal again, with the quieter signal partially removed. This results in 160 total signals plotted. While the pp-plot is somewhat closer to the diagonal, it still shows inconsistent parameter recovery, with some injections recovered completely outside the posterior. The combined p-value is still essentially 0.

At the bottom, we show the result for recovery with JPE, where our data template properly models both mergers present in the strain. As we recover both signals, we plot 160 signals total. We see that the fraction of events in which we completely missed the injection in the recovery properly goes to 0. The p-values are much higher than for the previous methods with the lowest p-value again obtained for $a_1$ at 0.085. The combined p-value is 0.337, indicating that posteriors obtained from JPE are self-consistent.

We conclude that our extension of RB to overlapping signals works and we can recover accurate posteriors with JPE. Hierarchical subtraction proved unreliable. Posteriors recovered with it are inconsistent, even as they do improve with each additional PE step.

\section{\label{sec:bias_estimation}Estimation of bias due to overlapping signals}
Running JPE on all pairs of potentially overlapping signals should lead to accurate posteriors as all the signals would be properly modeled in the likelihood. However, it is computationally inadvisable, as the method is much slower than the standard PE---in this paper we find that JPE on two signals is 6 times slower than SSPE. It is therefore of interest to develop a method that could ascertain if JPE is needed for a given pair of signals.

To test different methods of estimating PE bias due to overlapping signals, we generate an astrophysical population of BBHs and inject the signals into the detector network described in Sec.~\ref{sec:jpe_test} into zero noise, to focus on effects coming from the overlaps.
We draw the component masses from the Power-law + Peak distribution from~\cite{LIGOScientific:2021psn} and the component spins form the \textsc{Default} spin model from the same paper. We draw the redshifts from the merger rate reconstructed in Oguri's fit~\cite{Oguri2018EffectMergers}. All the other parameters are drawn from isotropic distributions, except time of coalescence which we set to 0 for all the events.~\footnote{This guarantees a large fraction of events for which overlapping signals affect PE.} We draw 1000 superthreshold events ($\mathrm{SNR} > 8$) from our population and arrange them into 500 pairs of overlapping signals.

\subsection{Prior-informed Fisher matrices}
Fisher-based approaches are common for studies into the effects of overlapping signals in the 3G era~\cite{Antonelli:2021vwg,johnson2024}. We incorporate the prior information into them and compare how well their predictions hold up against the PE posteriors.

The Fisher matrix mathod relies on applying the linear signal approximation~\cite{Finn:1992wt} to expand the waveform model $h(\vec{\theta})$ near the maximum likelihood point $\vec{\theta}_{\mathrm{ML}}$:

\begin{equation}\label{eq:lsa}
    h(\vec{\theta}) \approx h(\vec{\theta}_{\mathrm{ML}}) + \partial_i h(\vec{\theta}_{\mathrm{ML}}) \delta \theta^i \, .
\end{equation}

For the case of two signals present in the data, with just one modeled by the waveform, the approximated likelihood takes form:
\begin{equation}\label{eq:fisher_likelihood}
    \log\mathcal{L}(d | \vec{\theta})  \approx -\frac{1}{2} (\theta^i-\theta^i_{\mathrm{ML}}) \Gamma_{ij} (\theta^j-\theta^j_{\mathrm{ML}})\, ,
\end{equation}
where $\Gamma_{ij} = \langle \partial_i h | \partial_j h \rangle$ is the Fisher information matrix and the maximum likelihood point is located at:
\begin{equation}
    \theta_{\mathrm{ML}}^i =  \theta_A^i + \Gamma^{-1}_{ij} \langle \partial^j h | n \rangle + \Gamma^{-1}_{ij} \langle \partial^j h | h_B \rangle  \label{eq:likelihood_max}\, .
\end{equation}
Here $\Gamma^{-1}_{ij}$ is the covariance matrix, which is the inverse of the information matrix. We can interpret the second term on the right-hand side as shifting the likelihood peak due to the presence of noise in the data and the third term as shifting the peak due to bias introduced by the presence of the second signal~\cite{Cutler:2007mi}. For the details about the derivation one can refer to~\citet{Antonelli:2021vwg}, for example.

With the above approximation, we can predict the bias on parameter recovery we would expect from the presence of overlapping signals without performing expensive parameter estimation. According to the expression for the likelihood derived above, we would expect the posterior of parameter $\theta^i$ to be a Gaussian centered at $\theta^i_{\mathrm{ML}}$ with standard deviation $\sigma = \sqrt{\Gamma_{ii}^{-1}}$ (no summation is implied here by the use of repeated indices).

The location of the maximum likelihood point is determined up to the linear order and so breaks down as the bias due to the overlapping signal increases. We account for this by running a differential evolution algorithm~\cite{Storn1997} to refine our localization of the maximum. We use the Fisher-predicted value as our initial guess and set up a bounding box based on predicted parameter standard deviations. This helps to ensure that the likelihood expansion is performed near the true maximum and so improves the accuracy of the method. We use the implementation from the \textsc{SciPy} package~\cite{2020SciPy-NMeth}.

The standard Fisher-based approach ignores any priors and estimates the posteriors just from the likelihood. We follow~\citet{Dupletsa:2024gfl} in incorporating the prior information into the Fisher formalism. We sample from the Fisher likelihood truncated by the prior boundaries~\cite{Botev:2016}\footnote{We use the implementation from \url{https://github.com/brunzema/truncated-mvn-sampler}.}, and assign to each sample a weight equal to its prior probability $w_i(\vec{\theta}) = p(\vec{\theta})$. The resulting weighted samples behave as if they were drawn from  $\mathcal{L}(d | \vec{\theta})p(\vec{\theta})$ distribution instead of $\mathcal{L}(d | \vec{\theta})$ distribution they were drawn from.

\begin{figure}
    \centering
    \includegraphics[width=0.9\columnwidth]{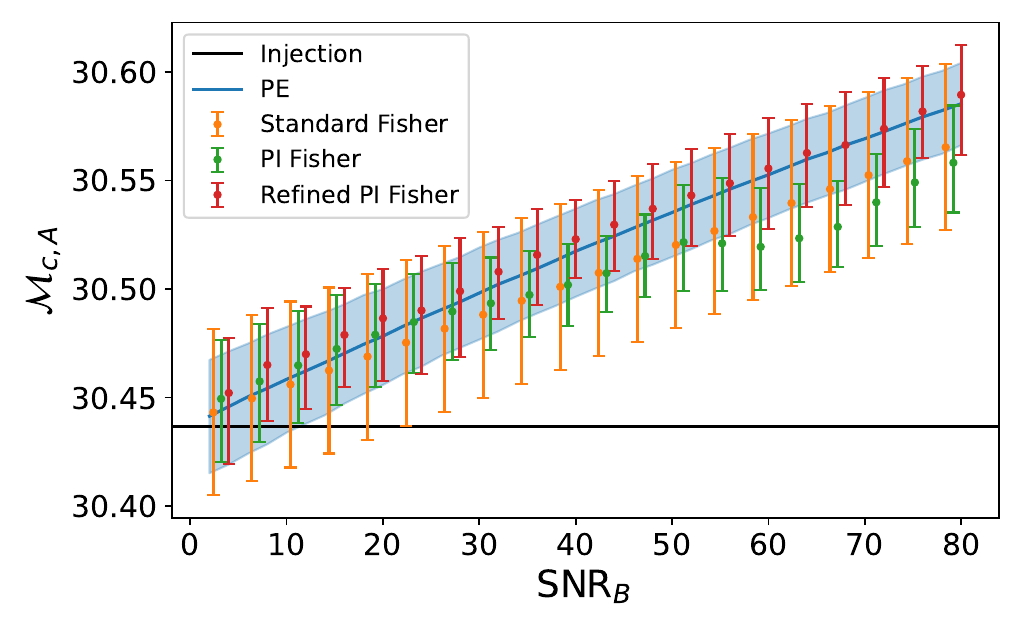}
    \caption{Estimation of the chirp mass measurements of the dominant signal for different bias estimation methods, as a function of SNR of the subdominant signal. The SNR of the dominant signal is fixed to 80. The errorbars and the color band indicate 90\% credible interval. The true PE posterior (blue) shifts linearly from the injected value of $\mathcal{M}_{c, A}$ (black) for all values of $\mathrm{SNR}_B$. The standard Fisher method (orange) predicts a posterior that is not shifted enough from the injected value and that is too wide. The prior-informed (PI) Fisher method (green) correctly predicts the posteriors for small $\mathrm{SNR}_B$, but breaks down as the subdominant signal gets louder. Refined PI Fisher method (red) numerically finds the maximum the maximum likelihood point and is similar to the PE posterior even for high SNR.}
    \label{fig:chirp_bias_fisher}
\end{figure}

We can see the effect of the improvements above in Fig.~\ref{fig:chirp_bias_fisher}. 
The standard Fisher method underestimates the shift in the posterior and overestimates its width. 
Incorporating the prior information makes the Fisher posterior much closer to the PE posterior, 
as long as the $\mathrm{SNR}_B$ of the quieter signal B is not too big. If we refine the location 
of the likelihood maximum with differential evolution, the Fisher method no longer fails for 
high $\mathrm{SNR}_B$. 

The above comparison was performed for a single draw of the signal parameters. While including the prior information leads to more accurate posteriors estimated with Fisher matrices, it is not a reliable method for all injections. In the following section, we show that the posteriors obtained with this method still differ from the PE posteriors and in some cases do not overlap with them at all.

\subsubsection{Test of the Fisher matrix method}

In this section, we asses our best performing Fisher method---prior-informed Fisher
with the maximum likelihood point refined by sampling.

\begin{figure}
    \centering
    \includegraphics[width=0.9\columnwidth]{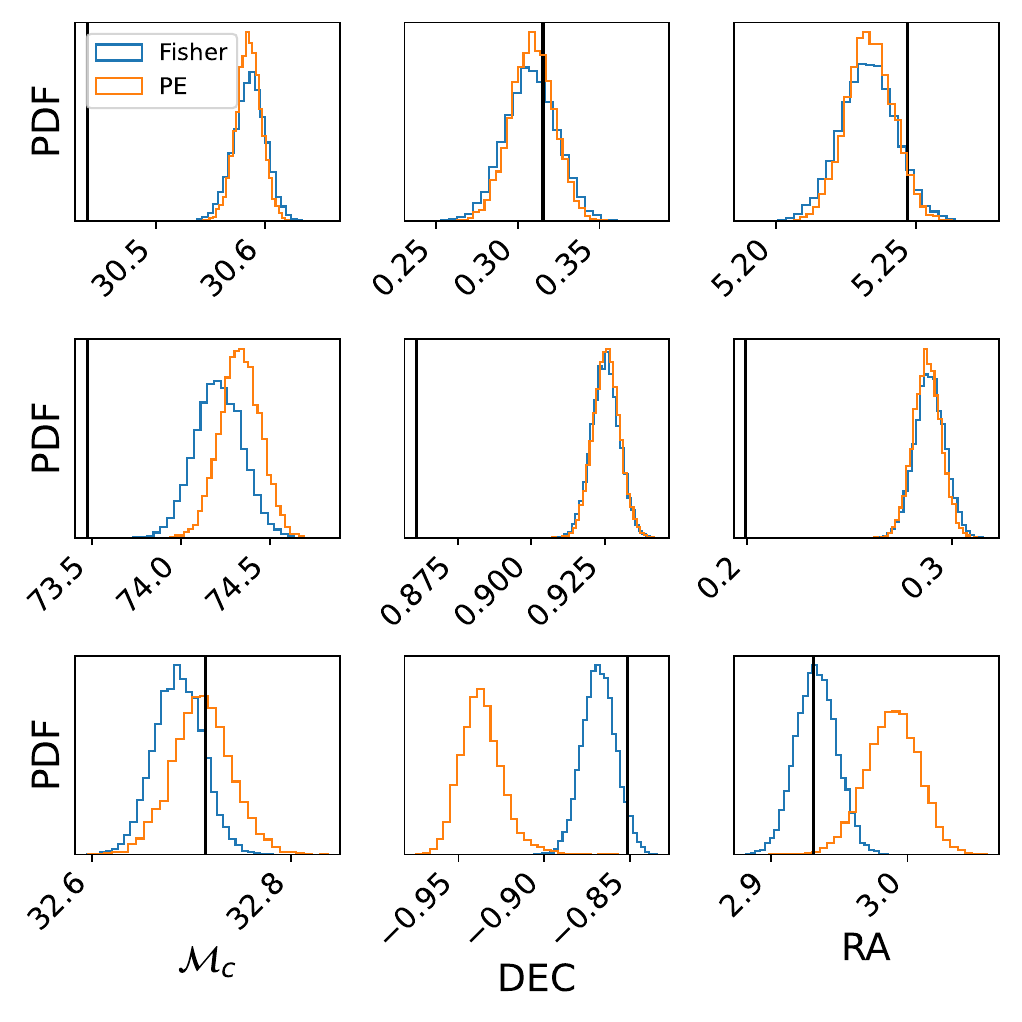}
    \caption{
        Comparison of the prior-informed Fisher posteriors (blue) with the PE posteriors (orange) and the injected value (black) for overlapping signals under the hypothesis of one signal present in the interferometer. \emph{Top:} example event for which the method works well. The significant shift from injection in chirp mass $\mathcal{M}_{c}$ is well predicted. \emph{Middle:} Example event for which the Fisher method succesfully predicts significant shift in sky localization. \emph{Bottom:} Example event for which the Fisher method fails to approximate PE posteriors, which is shifted away from the injection due to overlaping signals.
    } 
    \label{fig:fisher posteriors}
\end{figure}

We can see example 1D posterior for chirp mass ($\mathcal{M}_{c}$), declination (DEC), and right 
ascension (RA) in Fig.~\ref{fig:fisher posteriors}. These posteriors are also representative of other unimodal parameters, except the mass ratio that the Fisher method never recovers accurately.
In the top row, we see an example of the prior-informed Fisher matrix method (including numerical estimation of the maximum likelihood point) when it performs well. For most injections, it gives posteriors closely matching PE, even when the posterior is significantly shifted from the injection, like in the case of $\mathcal{M}_{c}$.
In the middle row, we show an example of a posterior where neglecting overlapping signals in the analysis leads to a large error in the sky localization. The Fisher method in general predicts this shift in posterior well, even when its predictions of $\mathcal{M}_{c}$ start to diverge from PE results.
Still, there are cases where the method breaks down, as in the example in the bottom row (we could detect no pattern which parts of the parameter space are prone to this). In particular, the DEC posterior is completely separated from the PE posterior. If we used the Fisher method to predict the bias due to overlapping signals in this case, we would be completely off. We would underestimate the bias due to overlaps and conclude the posteriors are much less affected than they actually are.

To summarize how well the Fisher method performs among the all 500 injections, we quantify bias in each parameter by a single number:
\begin{equation}\label{eq:bias_def}
    \Sigma_{\theta} = |\bar{\theta} - \theta_{inj}|/\sigma \,,
\end{equation}
where $\theta$ is a parameter of the binary, $\theta_{inj}$ is the injected value of the parameter, $\bar{\theta}$ is the mean of the posterior and $\sigma$ its standard deviation. It quantifies the shift in the posterior relative to its width. We take the absolute value for the convenience of comparison across multiple orders of magnitude. This is the same definition of bias used in~\citet{Antonelli:2021vwg} and measures how far the mean of the posterior is located away from the injection. It is not particularly useful for parameters with multiple peaks, so we only use it for parameters with unimodal posteriors.

\begin{figure}
    \centering
    \includegraphics[width=0.9\columnwidth]{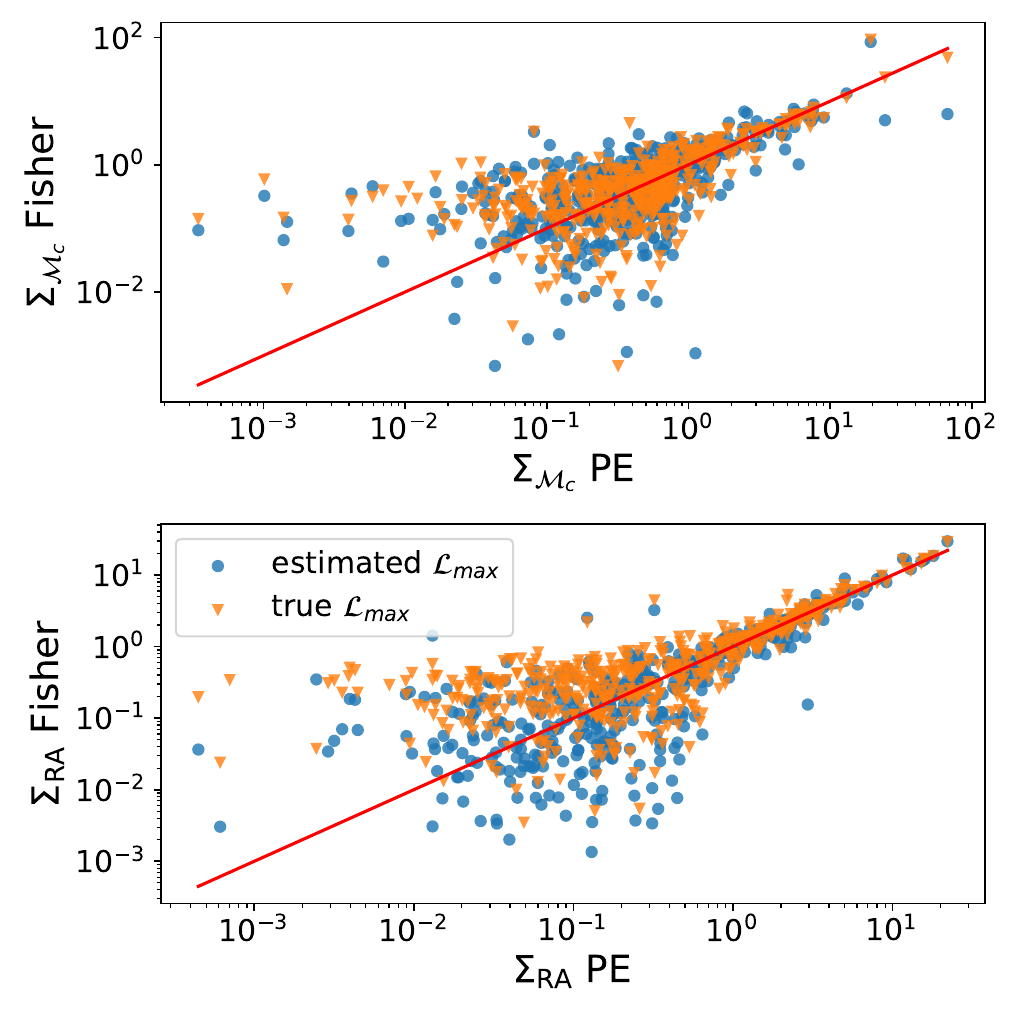}
    \caption{Bias $\Sigma_{\theta} = |\bar{\theta} - \theta_{inj}|/\sigma$ on the parameter estimation due to the presence of another signal, not accounted for in the template. The bias from PE is compared to the bias predicted with Fisher matrix method. The results are shown for chirp mass $\mathcal{M}_c$ and right ascension RA, representative of other parameters. Two different methods of estimating Fisher bias are plotted. One estimates the maximum likelihood point described and the other uses the true maximum likelihood point obtained from PE.}
    \label{fig:bias_fisher_maxl}
\end{figure}

In Fig.~\ref{fig:bias_fisher_maxl} we show how the bias estimated from the Fisher matrix compares to the bias obtained from PE. We chose to highlight the bias on chirp mass $\mathcal{M}_c$ and right ascension, as their behavior is representative of other parameters. We plot the results of Fisher expansion around the maximum likelihood point predicted with Fisher formalism and the location of the true maximum obtained from the sampling. The latter serves as an upper bound on accuracy achievable by Fisher matrix.

Above $\Sigma\approx1$ we see that the predicted and the true bias cluster close to the diagonal line, 
indicating they agree rather well. 
Note however the logarithmic nature of the plot; there can still be a factor of 2 difference between them. 
This is due to the sensitivity of our definition of bias $\Sigma_{\theta}$ to the standard deviation 
of the posterior for high biases. For example, the posteriors in the upper left corner of 
Fig.~\ref{fig:fisher posteriors} have a difference of $\Sigma_{\mathcal{M}_c}$ of 2.7 even though 
the predicted absolute shift is the same. Most outlier points in this region get close to the 
diagonal after using the true maximum likelihood point, indicating that the Fisher approximation 
used for finding this point starts breaking down for biases far away from the injection.

For $\Sigma\lesssim1$, so for the bias at or below the effects of the detector noise, the Fisher matrices fail to accurately predict the bias. The Fisher method can predict a relatively substantial bias of $\Sigma\approx1$ even for small true bias in the $0.001-0.1$ range. As such it is not accurate enough to serve as a predictor of whether we need to perform JPE or just a standard PE. The standard deviation of the mean scales as $\sigma/\sqrt{N}$ where $\sigma$ is the standard deviation of the distribution and $N$ is the number of drawn samples. With $N$ in the $2000-10000$ range, the uncertainty in our definition of bias due to sampling is $\sim0.01-0.02$. We would need the Fisher approach to be two orders of magnitude more accurate to predict small biases. These issues persist even when we expand near the true maximum likelihood point so we find Fisher formalism only useful when it predicts biases above the noise level $\Sigma\approx1$.

\subsection{Time-frequency overlap}
Previous work~\cite{johnson2024} with Fisher matrices indicated that the biases due to overlapping signals occur when time-frequency tracks of the signals intersect, which is equivalent to nonvanishing cross-correlation between the signals. We verify this explicitly with the PE results on our 500 astrophysical injections into zero noise. We plot the Jensen-Shannon divergence (JSD)~\cite{lin1991} between posteriors obtained with one and two signals in the strain as a function of the match $|M|$ between the signals $A$ and $B$:
\begin{equation}
    M = \langle h_A| h_B \rangle /\sqrt{\langle h_A| h_A \rangle\langle h_B| h_B \rangle} \,,
\end{equation}
which is just normalized cross-correlation. JSD is a measure of the statistical distance between distributions $p$ and $q$. It is defined by
\begin{equation}
    \mathrm{JSD}(p,q)=\frac{1}{2}D(p||m) + \frac{1}{2}D(q||m)\,,
\end{equation}
where $m=(p+q)/2$ is the mixture distribution and
\begin{equation}
    D(p||q) = \int\mathrm{d}x\, p(x)\ln\frac{p(x)}{q(x)}
\end{equation}
is the relative entropy~\cite{kullback1951} between the distributions. We chose JSD as our distance measure as it is symmetric between the distributions. In ~\citet{shaw2020} the authors have shown that when sampling with \textsc{Bilby} package statistical fluctuations of the sampler lead to JSD of up to 0.0015 nats between the marginalized 1D posteriors. We adopt $\mbox{JSD}>0.0015$ as our criterion of bias due to presence of overlapping signals.

\begin{figure}
    \centering
    \includegraphics[width=0.9\columnwidth]{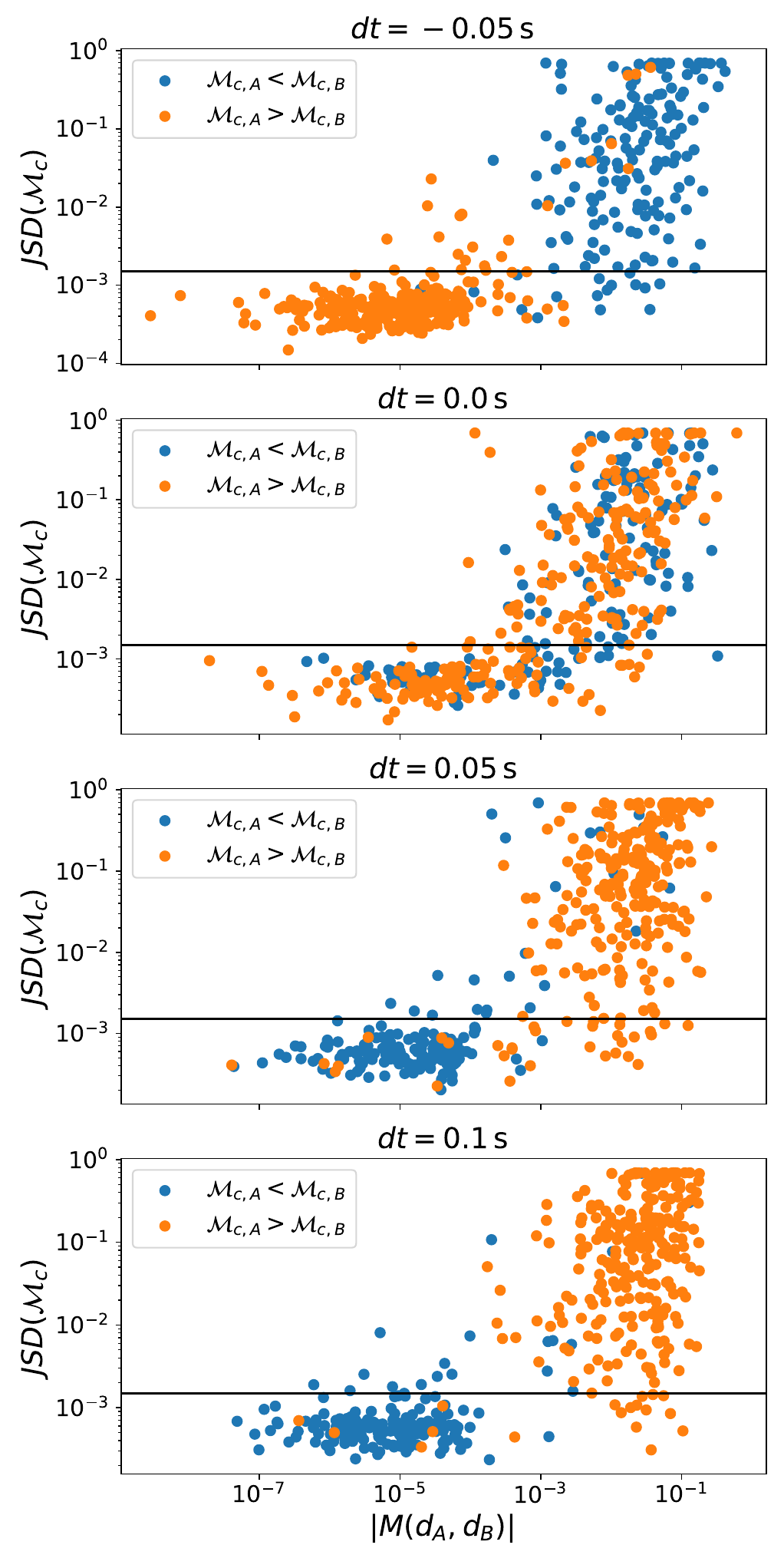}
    \caption{JSD between chirp mass posteriors with and without overlapping signals as a function of the match between the signals for different time separation of the mergers $dt=t_B-t_A$ at geocenter. As soon as the signals are separated temporarily, the chirp mass ordering starts predicting bias due to overlaps ($\mathcal{M}_{c,A}>\mathcal{M}_{c,B}$ mergers are expected to overlap in time-frequency for $dt>0$).
    The horizontal line at 0.0015 nats indicates JSD expected due to statistical fluctuations of the sampler.}
    \label{fig:jsd_match}
\end{figure}

The results can be seen in Fig.~\ref{fig:jsd_match}. We plot the JSD between the chirp mass posteriors obtained through PE with one and two signals in the strain as a function of the match between the signals. We see that more similar signals (higher match) lead to larger distortion of the posteriors due to overlaps (higher JSD) as predicted.

We also plot the same results where we shifted the separation of the signals and kept all the other parameters the same, and we see an interesting pattern emerge. At $dt=0\,\mathrm{s}$ both signals are coincident at geocenter but it does not mean that they are coincident at the detectors. Depending on the sky positions of the signals, the ordering of the signals at the detectors can be different at each of the detectors. But as soon as the time difference between the mergers is larger than $0.045\,\mathrm{s}$ (the Earth's diameter in seconds) the ordering of the signal at the interferometers is consistent between the interferometers and independent of sky location. The match and JSD are then driven primarily by ordering of the chirp masses of the signals, consistent with biases occurring due to overlaps in time-frequency. The chirp mass determines the leading order frequency evolution of the signal~\cite{2017AnP...52900209A}:
\begin{equation}
    \dot{f}=\frac{96}{5}\pi^{8/3}\mathcal{M}_c^{5/3}f^{11/3}\,.
\end{equation}
Up to leading order, the conditions $t_B > t_A$ and $\mathcal{M}_{c,A} <\mathcal{M}_{c,B}$ imply that the time-frequency tracks never cross, resulting in small match and small effect of overlaps on PE.

\begin{figure}
    \centering
    \includegraphics[width=0.9\columnwidth]{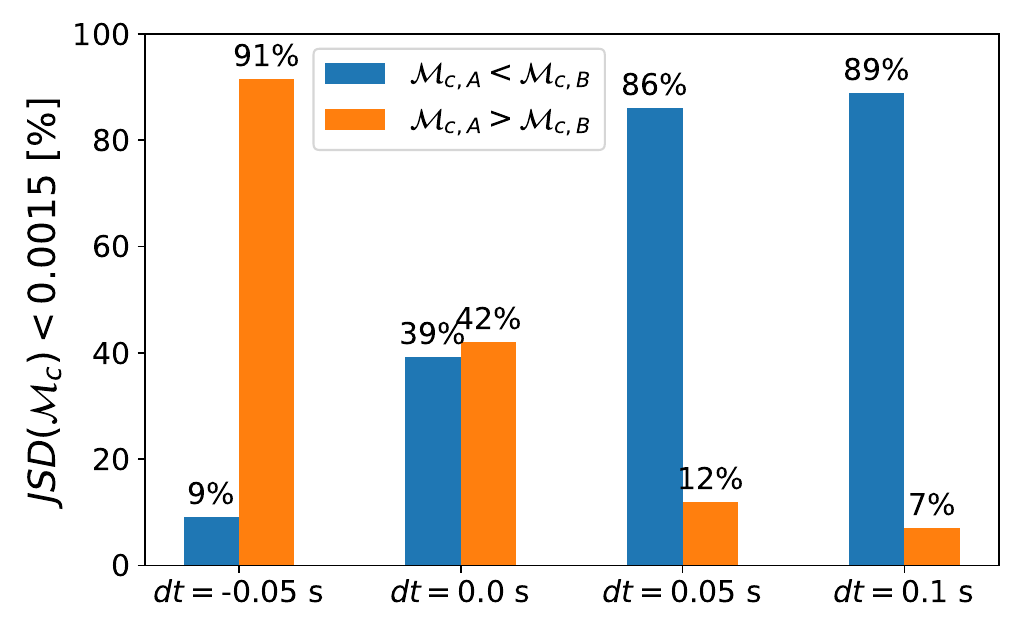}
    \caption{Percentage of injections for which presence of overlapping signals leaves chirp mass 
    posteriors unbiased ($\mbox{JSD}<0.0015$ compared with single injected signal posterior). For $dt>0.045\,\mathrm{s}$ chirp mass ordering can act as a fairly accurate indicator of bias.}
    \label{fig:jsd_hist}
\end{figure}

In Fig.~\ref{fig:jsd_hist} we demonstrate that the chirp mass ordering can act as a fairly accurate estimator of bias due to overlapping signals even for close mergers (as long as they are further apart than $0.045\,\mathrm{s}$). For $dt=0.05\,\mathrm{s}$ the chirp mass ordering leads to 86\% accurate classification of merger pairs with no bias and 88\% accurate classification of merger pairs with bias.

\subsection{Reweighting single-signal posteriors}
\begin{figure}
    \centering
    \includegraphics[width=0.9\columnwidth]{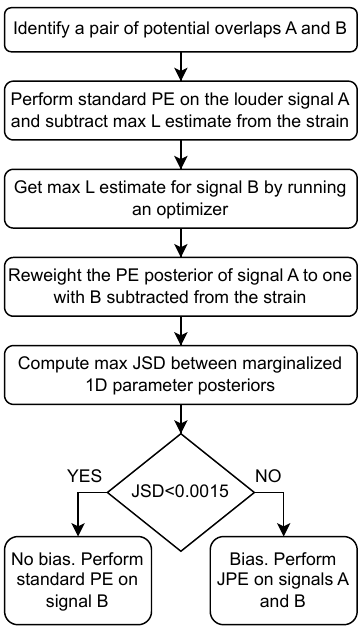}
    \caption{Schematic flowchart on how to determine if overlapping signals caused bias in PE.}
    \label{fig:bias_chart}
\end{figure}

Here we propose a novel method to determine biases due to overlapping signals based on the initial PE performed on the data. It does not rely on Fisher matrices and thus takes the full shape of the likelihood into account. In this paper, we focus on pairs of overlapping BBH signals, but the method is easily generalizable.

We start by going over the list of triggers from the search pipeline and identifying all the pairs of signals overlapping in time~\footnote{For the general deployment of this method. For the purpose of this paper, we assume all our injections triggered the search pipelines.}. Search pipelines will still be able to perform this task in the 3G era~\cite{relton2022}. We choose the higher SNR merger in the pair and perform the standard PE algorithm, completely ignoring the presence of overlapping signals---we use the likelihood $\mathcal{L}(\vec{\theta}_A|d)$. We then subtract the highest likelihood estimate of the signal $h(\hat{\theta}_A)$ from the strain. For the other event, we use the differential evolution algorithm~\cite{Storn1997} to find the maximum of $\mathcal{L}(\vec{\theta}_B|d-h(\hat{\theta}_A))$ and set it as our estimate of its parameters $\hat{\theta}_B$. The time needed to find the approximate maximum is negligible compared to the time needed for the sampling. We then use this estimate to reweight our posterior of signal A to one with signal B removed from the strain. Explicitly, we assign to each sample a weight:
\begin{equation}
    w_i = \frac{\mathcal{L}(\vec{\theta}_A^i|d-h(\hat{\theta}_B^i))}{\mathcal{L}(\vec{\theta}_A^i|d)}\,.
\end{equation}

If the reweighted posterior is the same as the original posterior (JSD $< 0.0015$ nats for all 1D 
posteriors), the presence of overlapping signals was not important for inference. We can perform 
standard PE on signal B and we spend the same amount of computational resources as if there was 
no possibility of overlaps. If the posteriors are different, we run the full JPE analysis on the pair. 
Compared with starting from JPE from the beginning, computational cost increases around 17\% by requiring 
the initial PE run. This does not pose a problem, as only a small fraction of mergers in the 3G era are 
expected to be affected by overlapping signals. It is also much lower than around 500\% increase in the 
computational cost if we run JPE when only standard PE run is needed.
In Fig.~\ref{fig:bias_chart}, we show a schematic flowchart of our proposed method.

\subsubsection{Test of the posterior reweighting method}

\begin{figure}
    \centering
    \includegraphics[width=0.9\columnwidth]{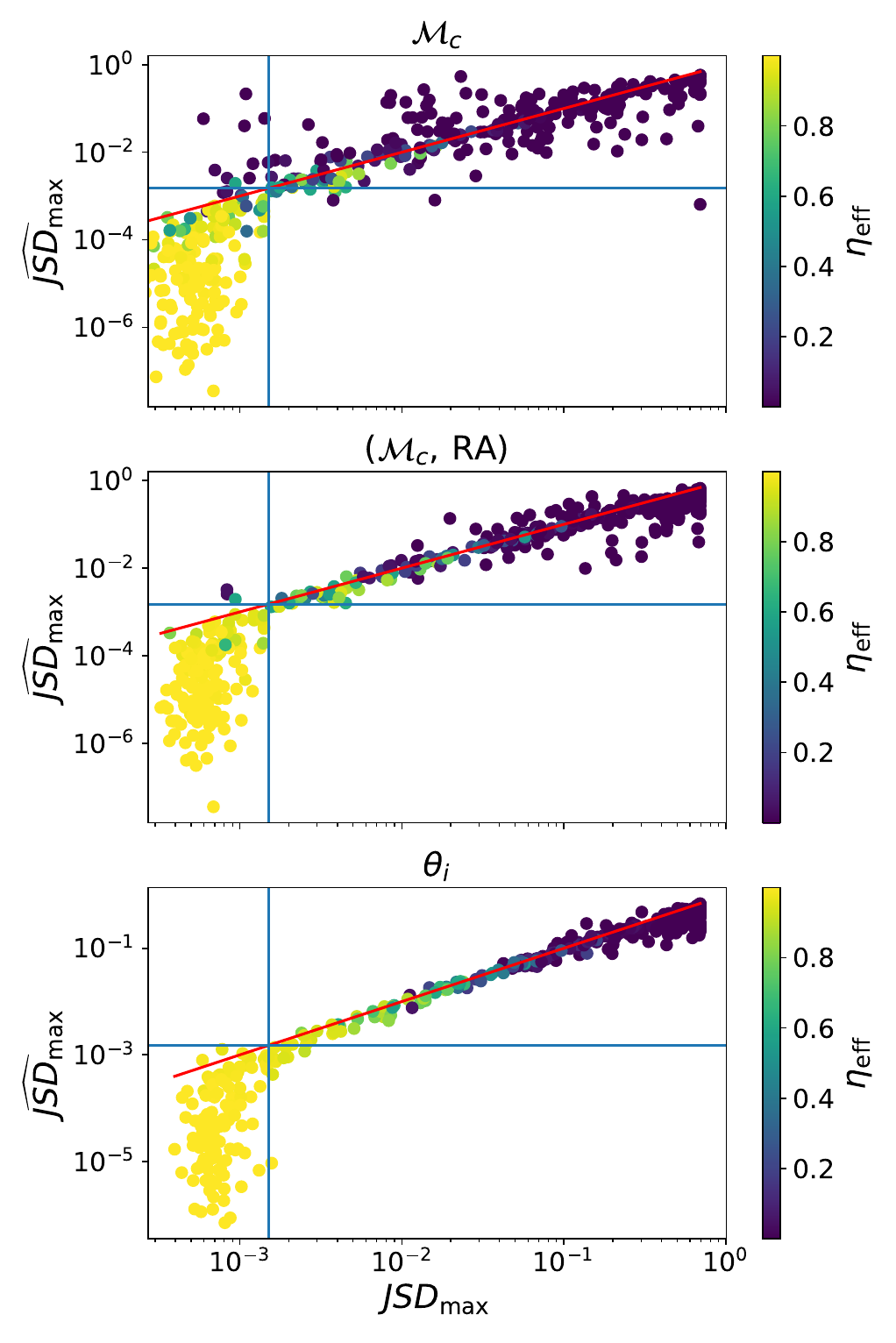}
    \caption{The predicted JSD ($\widehat{JSD}_\mathrm{max}$) between overlapping and non-overlapping signal posteriors plotted against the true JSD (${JSD}_\mathrm{max}$) for 500 injections into zero noise. For a single parameter \emph{(top)} it follows a clear trendline. As we take maximum JSD from more parameters \emph{(middle)}, the prediction gets better until we add all the sampling parameters \emph{(bottom)}. The points are colored according to the resampling rate $\eta_\mathrm{eff}$ to the reweighted posterior. The blue lines indicate a JSD of 0.0015 nats.}
    \label{fig:jsd_prediction}
\end{figure}

In Fig.~\ref{fig:jsd_prediction} we show how the JSD predicted by our method ($\widehat{JSD}_\mathrm{max}$) compares to the true JSD between the single-signal posterior and the overlapping signals posteriors (${JSD}_\mathrm{max}$) based on our 500 injections into zero noise. We see that reweighting PE posteriors leads to an accurate estimate of JSD between them, as long as we use the 1D posteriors of all our sampling parameters. It is less reliable if we use just a single parameter due to low reweighting efficiency in cases of significant bias. When the effects of overlaps shift the posteriors significantly in one parameter (high true JSD), the original posterior poorly covers the target posterior resulting in a low effective sample size. This leads to high predicted JSD for all parameters, even the ones that should be unaffected by the overlaps. If we use maximum predicted JSD from across all the parameters, this high predicted JSD will only happen when the true JSD is high as well and high precision is not important. This effect is responsible for higher scatter around the line of equal JSD for signal pairs with high JSD.

We see an interesting effect in areas of low true JSD, below the 0.0015 nats expected from statistical fluctuations during the sampling. The JSD is consistently underestimated in this region.  This shows us that for these mergers, the effect of accounting for overlaps is orders of magnitude lower than fluctuations due to sampling. When we reweight the posterior we are relying on the same realization of the sampler and do not get up to 0.0015 nats of JSD difference between the distributions we get from our PE run that is used for calculating the true JSD.

An important part of each figure is the lower-right quadrant created by the lines of $JSD=0.0015\,\mathrm{nats}$, which in the classification task represents wrongly concluding the overlaps cause no bias in PE when they do distort the shape of the posterior. There are only 5 points in this region indicating $\sim1\%$ false-negative rate for our method. For a science-conservative approach, we want the false-negative rate to be as close to zero as possible. To that end, we could move our JSD threshold to a lower number, like 0.0004 nats. This would be enough to accurately classify all but one cases. For a computation-conservative approach, we may decide that the 1\% false-negative rate is acceptable, especially as the true JSD of the points in this region is still low, below 0.003 nats. This would lead to a minimal bias in the recovered posteriors, see Fig.~\ref{fig:jsd_posteror_diff}.

\begin{figure}
    \centering
    \includegraphics[width=0.9\columnwidth]{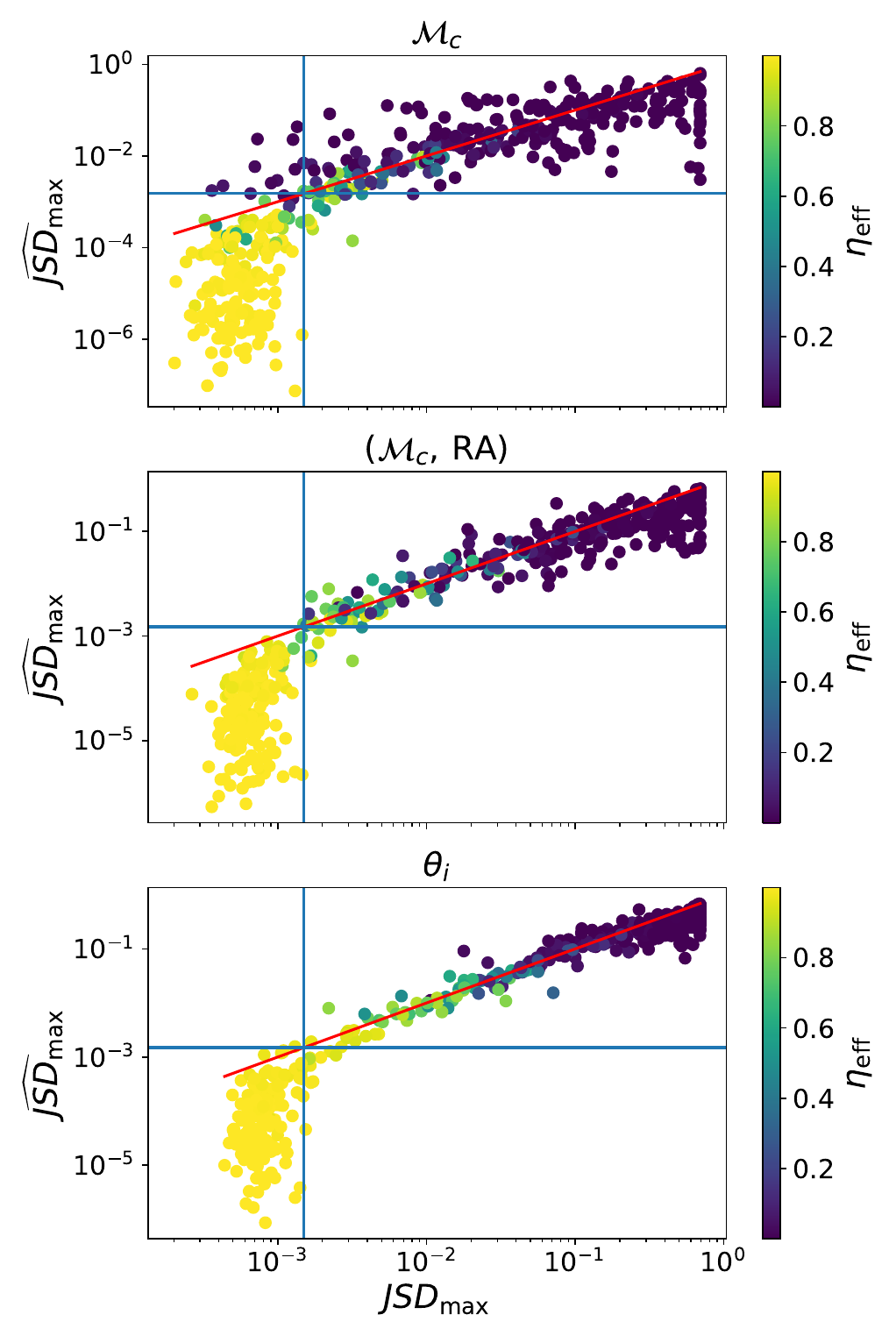}
    \caption{The predicted JSD ($\widehat{JSD}_\mathrm{max}$) between overlapping and non-overlapping signal posteriors plotted against true JSD (${JSD}_\mathrm{max}$) for 500 injections into Gaussian noise. For a single parameter \emph{(top)} it follows a clear trendline. As we take maximum JSD from more parameters \emph{(middle)}, the prediction gets better until we add all the sampling parameters \emph{(bottom)}. The points are colored according to the resampling rate $\eta_\mathrm{eff}$ to the reweighted posterior. The blue lines indicate a JSD of 0.0015 nats.}
    \label{fig:jsd_prediction_noise}
\end{figure}

The above discussion was based on our injection into zero-noise. In Fig.~\ref{fig:jsd_prediction_noise} we repeat the process for the same injections, but injected into Gaussian noise. It follows a similar pattern to the zero noise injections with a wider scatter around the line of accurately predicted JSD. As such, the accuracy of the method drops and the fals-negative rate rises to 2\%. Again, this can be lowered by changing our JSD cutoff at the cost of higher false-positive rate.

\section{\label{sec:conclusions}Conclusions}

Mergers of binary black holes with close coalescence times in the 3G era can cause significant biases in parameter estimation if the overlapping signals are unaccounted for. If the bias is present, hierarchical subtraction can shift the posteriors closer to their true value in most cases, but the resulting posteriors are still distorted, with HS failing the pp-test. We need to properly model all signals with joint parameter estimation to recover accurate posteriors.

We adapted joint parameter estimation to work with relative binning for \textsc{IMRPhenomXPHM} as well as waveform models without higher modes nor precession. For binary black holes, it produces reliable parameter estimation results in the timeframe of 11 days for a pair of overlapping signals on a 16-core \textsc{Intel Skylake Gold 6148} CPU (for average signal SNR of 53). This can be used in future injection studies into Einstein's Telescope and Cosmic Explorer to properly model the effects of overlaps on the recovered properties of the signals.

We have tested how reliable prior-informed Fisher matrices are for the estimation of bias due to overlaps and found them reliable only when they predict bias above the value expected from the detector noise. The overlap of time-frequency tracks on the other hand proved to be a good indicator of the presence of bias due to overlaps. Even for mergers with $dt=0.05\,\mathrm{s}$ it accurately predicted if bias was present in 86\% of cases.

Finally, we proposed a parameter estimation-based approach to determining if the presence of overlapping signals causes bias. By reweighting the posterior to one with other signals subtracted and quantifying the shift, we can determine if the bias is present with 99\% accuracy in zero noise (98\% accuracy in Gaussian noise). If one is willing to pay the price of a few false 
alarms, one can push this to 100\% identification of cases needing joint parameter estimation. The choice depends on whether one prefers a science-conservative approach, or a computing-conservative approach. This method does not rely on any simplifications to the likelihood inherent in the Fisher formalism and properly takes into account the realization of the strain in the detector, but is much slower, being PE-based. 
However, compared with the resources that are anyway needed to run parameter estimation on the signals, it does not 
introduce significant overhead.

In future work, our method will be tested on signals with higher temporal separations. Due to 
the computational constraints of running parameter estimation, we have focused on binary black hole 
mergers close together in time. This was to guarantee a population for which bias due to overlaps 
played a significant role so that different parameter estimation techniques could be tested, and our 
method of predicting bias had a wide range of biases to verify it.
We leave it for future work to test the method on binary neutron stars, overlaps of which are expected 
to be common in the 3G era.

\acknowledgments
We thank Qian Hu for useful comments on the manuscript.
T.B., H.N., A.S. and C.V.D.B are supported by the research program of the Netherlands Organisation for Scientific Research (NWO). 
T.D.\ TD acknowledges funding from the EU Horizon under ERC Starting Grant, no.\ SMArt-101076369. 
Views and opinions expressed are however those of the authors only and do not necessarily reflect those of 
the European Union or the European Research Council. Neither the European Union nor the granting authority can be held responsible for them.
The authors are grateful for computational resources provided by Cardiff University, and funded by STFC grants ST/I006285/1 and ST/V005618/1 supporting UK Involvement in the Operation of Advanced LIGO.
The plots were prepared with \textsc{Matplotlib}~\cite{Hunter:2007}. \textsc{NumPy}~\cite{harris2020array} and \textsc{SciPy}~\cite{2020SciPy-NMeth} were used during data analysis.

\appendix
\section{\label{app:code_release}Code release}
We made our Fisher matrix and Joint Parameter Estimation codes available at github~\footnote{\url{https://github.com/corvus314/jointRB}} together with examples showing their use. They were developed to work together with \textsc{Bilby}~\cite{Ashton:2018jfp,Romero-Shaw:2020owr} parameter estimation software.

While our JPE code was extensively tested on the case of 2 signals, we developed it to work with an arbitrary number of signals present. We do note that this may 
cause a noticeable decrease in performance.

JPE with with higher modes requires reimplementing some functions from \textsc{LALSimulation}~\cite{lalsuite} and as such our implementation is hardcoded for the \textsc{IMRPhenomXPHM} waveform~\cite{Pratten:2020ceb}. We also include a separate implementation for a general \textsc{LALSimulation} waveform model, as long as it does not include higher modes or precession. While not a focus of this paper, this second implementation can give a quick parameter estimation of overlapping 
binary neutron stars.

\section{\label{app:jsd}Additional details on bias estimation method}

\begin{figure}
    \centering
    \includegraphics[width=0.9\columnwidth]{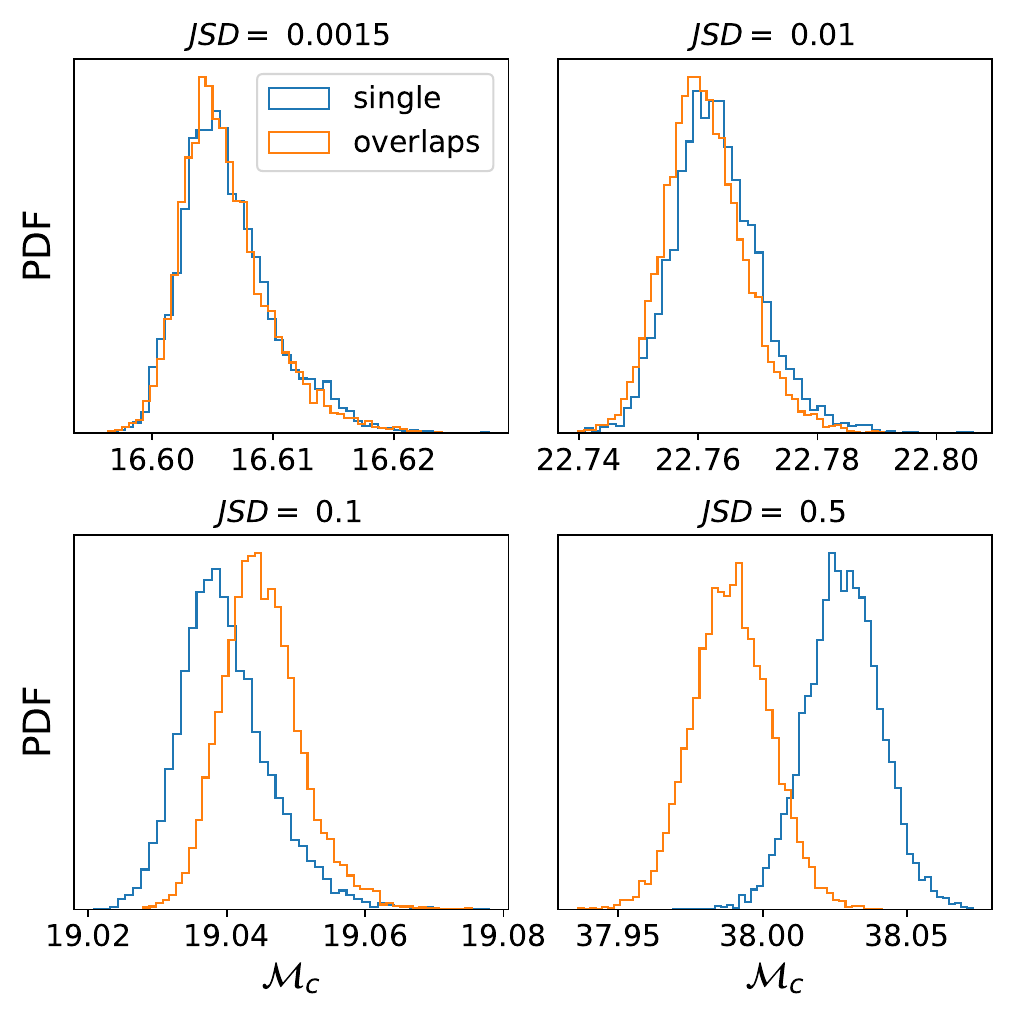}
    \caption{Example pairs of chirp mass posteriors with different levels of JSD between them.}
    \label{fig:jsd_posteror_diff}
\end{figure}

In this paper, we chose to quantify the bias due to overlapping signals with JSD between single-signal and overlapping posteriors. Compared with other measures, such as the shift of the parameter average from the injection, it is not an intuitive measure. As such in Fig.~\ref{fig:jsd_posteror_diff} we show examples of posterior pairs with different levels of JSD between them.

\begin{figure}
    \centering
    \includegraphics[width=0.9\columnwidth]{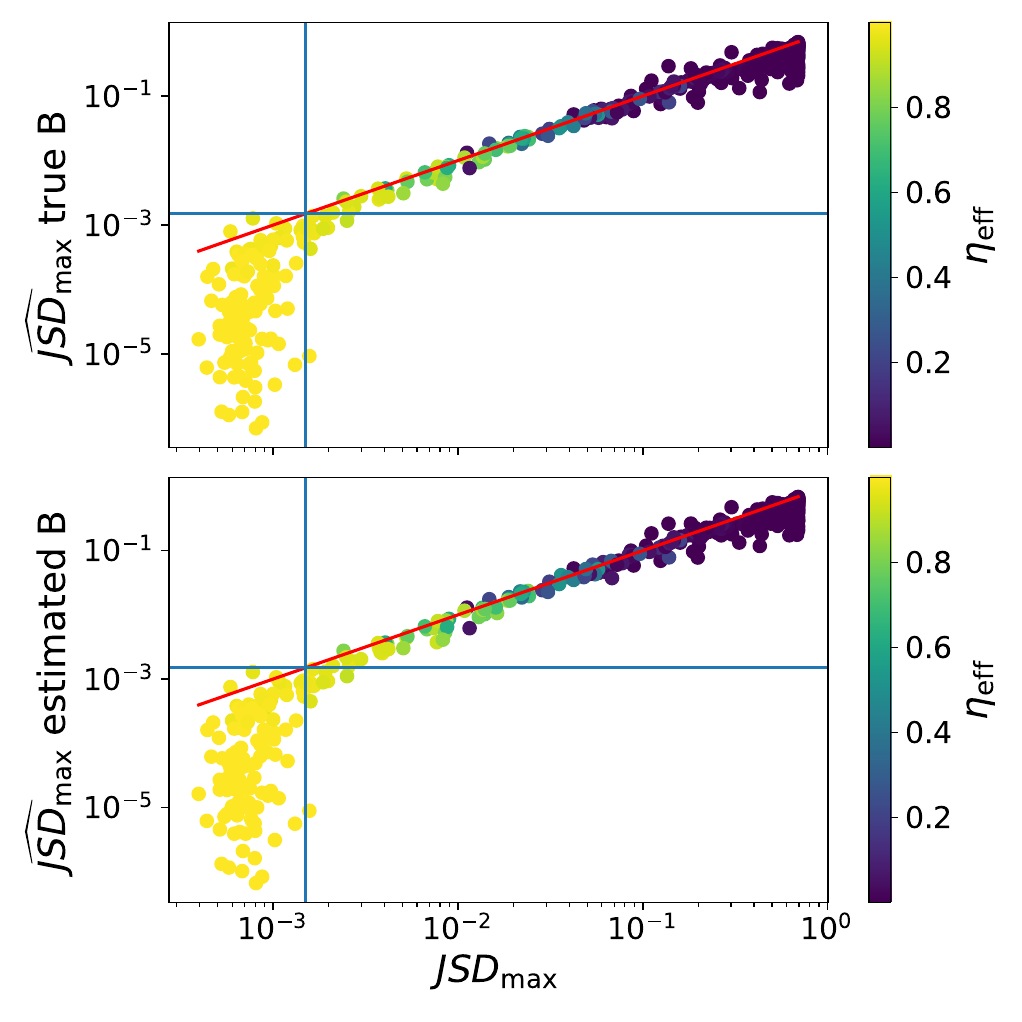}
    \caption{Performance of our bias estimation method when using the true parameters of signal B compared to using estimates of the maximum likelihood parameters. There is almost no difference in the posterior}
    \label{fig:jsd_estimated_true}
\end{figure}

In our method, we rely on performing PE just on a single signal and finding the best estimate of the maximum likelihood parameters of the other signal. This mismatch between the true and the estimated signal has only small effects on the performance of our method. We can see this in Fig.~\ref{fig:jsd_estimated_true}, where there is almost no difference between the two.

\begin{figure}
    \centering
    \includegraphics[width=0.9\columnwidth]{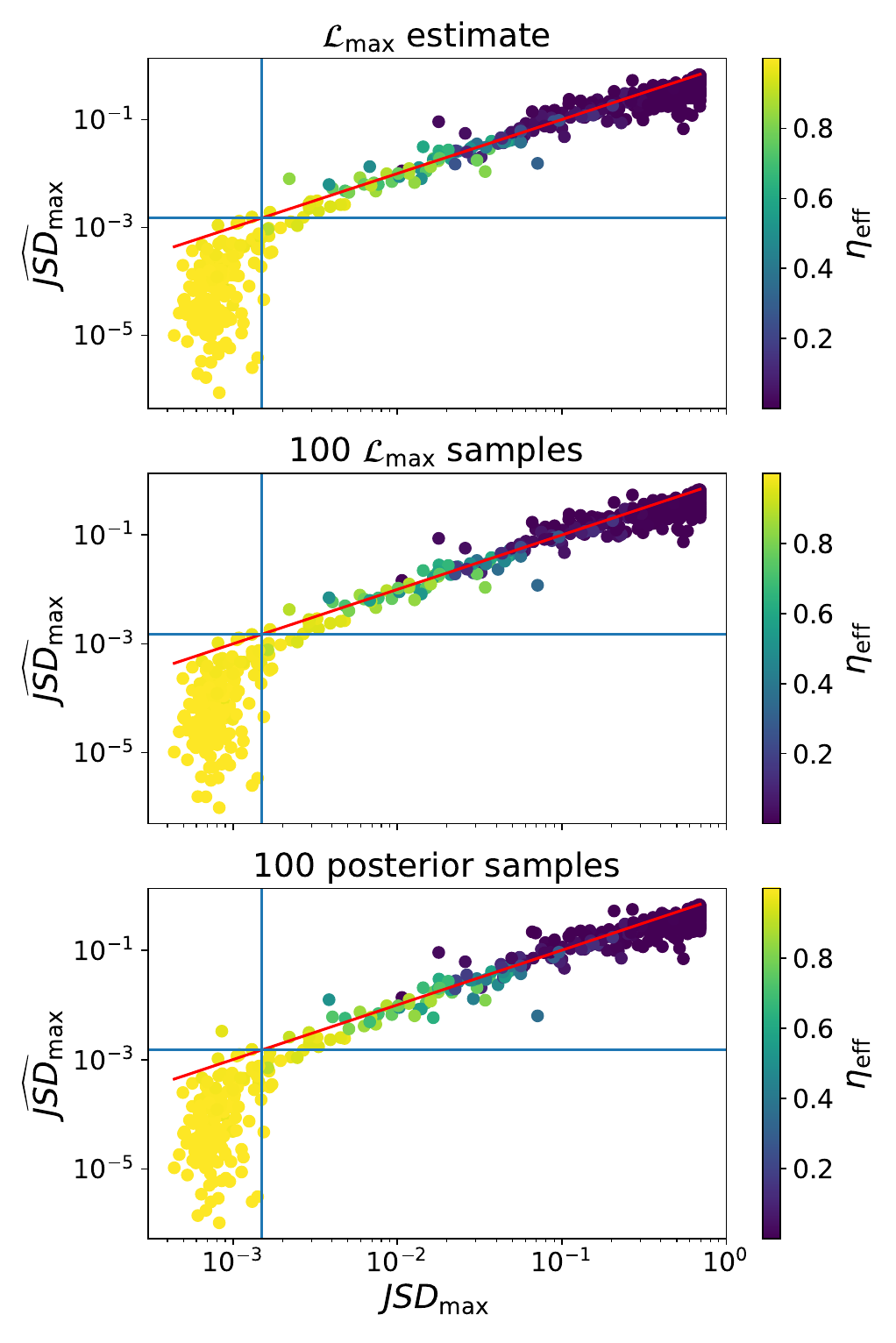}
    \caption{Adaptation of the bias estimation method for case where SSPE was performed on both overlapping signals. \emph{Top:}} Subdominant signal, is estimated by its maximum likelihood sample. \emph{Middle:} Reweighting is averaged over 100 samples with the highest likelihood from SSPE of the subdominant signal. \emph{Bottom:} Reweighting is averaged over 100 draws from the SSPE posterior of the subdominant signal.
    \label{fig:jsd_100_samples}
\end{figure}

If we were willing to perform initial PE on both signals instead of just one, our method could be improved further. Instead of reweighting the posterior of event A to one estimate of signal B, we could draw multiple samples, reweight the likelihood to each one, and then average it over the samples. If we are willing to pay the initial PE cost, this approach is still fast and parallelizable. We show the results in Fig.~\ref{fig:jsd_100_samples} based on taking 100 samples from the posterior of signal B. If we take the samples at random from the posterior, the method performs worse than with just the maximum likelihood estimate. If we take 100 samples with the highest likelihood, the method shows minor improvements.

\FloatBarrier

\bibliography{bibliography}

\end{document}